\documentclass[fleqn,usenatbib]{mnras}

\usepackage[T1]{fontenc}

\DeclareRobustCommand{\VAN}[3]{#2}
\let\VANthebibliography\thebibliography
\def\thebibliography{\DeclareRobustCommand{\VAN}[3]{##3}\VANthebibliography}

\usepackage{graphicx}	
\usepackage{amsmath}	
\usepackage{amssymb}	
\usepackage{xspace}	    
\usepackage{xcolor}	    

\usepackage{newtxtext,newtxmath}

\newcommand{\orcid}[1]{\href{https://orcid.org/#1}{\includegraphics[width=10pt]{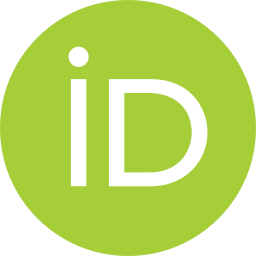}}}

\begingroup
\catcode`\_=\active
\gdef_#1{\ensuremath{\sb{\mathrm{#1}}}}
\endgroup
\mathcode`\_=\string"8000
\catcode`\_=12

\newcommand{\Msolar}{M$_{\odot}$\xspace} 

\newcommand{\HII}{H\textsc{ii}\xspace}

\newcommand{\rev}{\textcolor{black}}

\title[Bottling the Champagne]{Bottling the Champagne: Dynamics and Radiation Trapping of Wind-Driven Bubbles around Massive Stars}

\author[Geen et al]{
Sam Geen$^{1}$\thanks{E-mail: s.t.geen@uva.nl} \orcid{0000-0002-3150-2543},
Alex de Koter$^{1,2}$  \orcid{https://orcid.org/0000-0002-1198-3167}
\\
$^{1}$ Anton Pannekoek Institute for Astronomy, Universiteit van Amsterdam, Science Park 904, 1098 XH Amsterdam, The Netherlands\\
$^{2}$ Institute of Astronomy, KU Leuven, Celestijnenlaan 200D, 3001 Leuven, Belgium\\
}

\date{Accepted XXX. Received YYY; in original form ZZZ}

\pubyear{2021}

\begin{document}
\label{firstpage}
\pagerange{\pageref{firstpage}--\pageref{lastpage}}
\maketitle

\begin{abstract}
In this paper we make predictions for the behaviour of wind bubbles around young massive stars using analytic theory. We do this in order to determine why there is a discrepancy between theoretical models that predict that winds should play a secondary role to photoionisation in the dynamics of \HII regions, and observations of young \HII regions that seem to suggest a driving role for winds. \rev{In particular, regions such as M42 in Orion have neutral hydrogen shells, suggesting that the ionising radiation is trapped closer to the star.} We first derive formulae for wind bubble evolution in non-uniform density fields, focusing on singular isothermal sphere density fields with a power law index of -2. We find that a classical ``Weaver''-like expansion velocity becomes constant in such a density distribution. We then \rev{calculate the structure of the photoionised shell around such wind bubbles, and determine at what point the mass in the shell cannot absorb all of the ionising photons emitted by the star, causing an ``overflow'' of ionising radiation.} We also estimate perturbations from cooling, gravity, magnetic fields and instabilities, all of which we argue are secondary effects for the conditions studied here. Our wind-driven model provides a consistent explanation for the behaviour of M42 to within the errors given by observational studies. We find that in \rev{relatively denser} molecular cloud environments \rev{around single young stellar sources, champagne flows are unlikely until the wind shell breaks up due to turbulence or clumping in the cloud.}
\end{abstract}





\section{Introduction}
\label{introduction}

Ionising radiation from stars plays an important role in the interstellar medium. \rev{Lyman continuum} photons leave the star and \rev{photo}ionise the gas around them in a region bounded by an ionisation front\rev{, which separates this photoionised gas and the neutral gas outside.} These volumes of photoionised gas are called \HII regions. 

\rev{This photoionised gas recombines to neutral hydrogen over time, and hence a certain number of ionising photons are required to keep the gas ionised. The flux of photons reaching the ionisation front is also reduced by a \textit{geometric dilution factor}, which describes the fact photons leaving the surface of a star in radial directions are spread across increasingly large spherical shells. The ionisation front travels outwards if there are any remaining ionising photons reaching the front. Since the photoionised gas is warmer ($\sim10^4~$K) than the neutral gas outside (10-1000$~$K), this also causes thermal expansion of the \HII region. This expansion reduces the density of the photoionised gas, lowering its recombination rate and hence allowing ionising photons to reach larger radii before being absorbed, and hence the ionisation front moves outwards as the \HII region expands. The thermal expansion produces a leading shock wave that accelerates and condenses the surrounding gas into a dense shell of partially or wholly neutral gas. Models including these processes have been produced by, e.g., \cite{KahnF.D.1954}, \cite{ SpitzerLyman1978} and \cite{Dyson1980}.} 

\rev{In a uniform medium with densities similar to those found in star-forming molecular clouds, the behaviour of these solutions is typically for the ionisation front to move outwards rapidly over the \textit{recombination time} in the gas, until none of the ionising photons arrive at the ionisation front, and so the \HII region reaches \textit{photoionisation equilibrium}, where the total recombination rate balances the ionising photon emission rate from the star. The recombination time of dense gas inside molecular clouds is typically very short compared to the lifetime of massive stars that produce significant quantities of ionising radiation. After photoionisation equilibrium is achieved, the \HII region grows primarily due to thermal expansion at speeds slower than the sound speed in the photoionised gas.}

\rev{However, if the ionisation front moves into regions with lower gas densities, such as towards the outskirts of a molecular cloud, the outward movement of the ionisation front can accelerate since the more diffuse gas is less efficient at absorbing the ionising photons than the denser cloud material. In some cases, the \HII region leaves photoionisation equilibrium and the ionisation front travels outwards supersonically.} This mode of \HII region \rev{evolution} has been referred to as a  ``champagne'' flow by \cite{TenorioTagle1979}, who invoke a step function in density to model a flow leaving a dense cloud and entering the diffuse medium outside.

\rev{\cite{Comeron1997} produces analytic descriptions and 2D numerical simulations of such a case for a pair of massive stars embedded in the denser part of the step function, including both photoionisation and stellar winds. They find that winds produce complex structures as the wind bubble enters the diffuse medium, but that their presence does not significantly affect the evolution of a rapid champagne flow driven by photoionisation into the diffuse medium.}

Alternatively, \cite{Franco1990} show that a champagne flow occurs if the density gradient in the neutral cloud \rev{has a power law index steeper than} $-3/2$, (i.e. one in which the density halves when the radius increases by a factor of roughly 1.6). \rev{In this mode, the \HII region never reaches photoionisation equilibrium}. Simulations of star formation predict that stars are born in density peaks in clouds with a power law index around $-2$ \citep[e.g.][]{Bate2012,Lee2018}, and so are likely to form champagne flows as ionising radiation escapes the protostellar environment around the young star.

Simulations of feedback in molecular clouds with self-consistent star formation frequently reproduce such champagne flows. For example, \cite{Dale2012}, \cite{Ali2018} and \cite{Zamora-Aviles2019} include ionising radiation in their simulations and report champagne flows resulting from such conditions. \cite{Ali2018} target the Orion nebula, and find some similarities with the observed region, including far ultra-violet (FUV) emission and expansion of gas above 10 km/s. Simulations including winds and photoionising radiation from massive stars, such as \cite{Dale2014} and \cite{Geen2020} also report similar flows, with winds embedded inside them. \cite{Geen2015b} find that magnetic fields can constrain the fragmentation of the shells around \HII regions and hence the escape of ionising radiation to some extent, although even with magnetic fields, ionising radiation and warm ($\sim 10^4~$K) gas drives rapid expansion of champagne-like flows into the external medium.

However, observers such as \cite{Guedel2008} and \cite{Pabst2019} find evidence that nearby \HII regions such as the Orion Nebula are filled with x-ray emitting gas surrounded by a rapidly-expanding neutral shell. This is characteristic for stellar winds, which are emitted from the star at thousands of km/s \citep{Lamers1993,Vink2011} and subsequently shock-heats the gas around the star to millions of degrees \citep[e.g.][]{Weaver1977,Dunne2003}. Since the shell around this wind bubble is neutral, it implies that any ionising radiation from the star is trapped by the stellar wind bubble. The hypothesis is that the stars in such regions are not producing champagne flows driven by ionising radiation despite having the conditions for such flows according to the previously described theoretical models.

The purpose of this work is to use algebraic prescriptions to identify how such systems should behave and provide a stepping stone between observed behaviour of feedback structures around young massive stars, and more detailed numerical models. We apply analytic theory to power law density fields where champagne flows can occur \citep[e.g.][]{Franco1990}. \rev{We first give an overview of the physical systems modelled. We then derive a classical ``Weaver-like'' solution \citep{Weaver1977} for an expanding wind bubble in power law density fields. In the next Section, we introduce semi-analytic and algebraic solutions for the evolution of the photoionised shell around the wind bubble and the point at which ionising photons can overflow the neutral shell. We then discuss possible perturbations to this model, and compare our results to M42 in Orion.}

\rev{The key argument of this paper is that there exists a set of solutions for a wind-driven bubble expanding into a power law density field where the ionising radiation remains trapped, preventing a ``champagne'' flow as described by \cite{Franco1990}. Mathematically, there is a point where this trapping should end. For typical conditions in molecular clouds around individual massive stars, our model predicts that any possible champagne flow is likely to remain ``bottled'' by the shell around the wind bubble for the duration of the model's applicability, unless the cloud is sufficiently diffuse. Other 3D environmental effects are more likely to allow ionising radiation to break out from the cloud, such as nearby wind bubbles merging \citep[e.g.][]{Calderon2020}, the density profile in the cloud flattening out as the wind expands into the wider cloud environment, or a step function to a much lower density outside the cloud as described by \cite{TenorioTagle1979}. }



For the rest of this Section we review the literature regarding the modelling of photoionised \HII regions with embedded wind bubbles, and lay out the structure of the rest of this paper.

\subsection{Theory of Stellar Wind Feedback}

Early theoretical work by \cite{Avedisova1972}, \cite{Castor1975} and \cite{Weaver1977} established a basis for understanding adiabatic wind bubbles, as well as initial work on the behaviour of the bubbles in response to radiative cooling and ionising radiation. 

More recent work by \cite{Capriotti2001} looked in more detail at the dynamical evolution of wind bubbles inside photoionised \HII regions, and argued that wind bubbles play a secondary role to photoionisation. \cite{Haid2018} found similar results in controlled hydrodynamic simulations in a uniform medium. This was also the conclusion of \cite{Geen2019} in a power law density field where an efficiently cooling wind bubble was invoked inside an existing photoionised \HII region.

However, in conditions where wind bubbles retain most of their energy, they can efficiently drive expansion of hot ($>~10^6~$K) bubbles around massive stars. Whether or not wind bubbles retain most or all of the energy deposited by the star is an open question. \cite{Fierlinger2016} argue that properly resolving the contact discontinuity between the hot, diffuse wind bubble and the denser shell around it is critical in setting the energetics of the wind bubble. \cite{Gentry2016} argue that the same is true for supernovae, and that numerical diffusion in grid hydrodynamic codes can cause artificial energy losses. \cite{Fielding2020} and \cite{Tan2021} argue that turbulent mixing can increase the cooling rate of the wind bubble, provided the turbulence is strong enough. Increased wind pressure also increases the density of the shell around the wind bubble, more effectively trapping the ionising radiation.

The role of radiation pressure in \HII regions has been explored in models by \cite{Mathews1967}, \cite{Krumholz2009}, \cite{Murray2010}, \cite{Draine2011} and \cite{Kim2016}, both with and without an embedded wind bubble. In the model of \cite{Pellegrini2007}, which includes an extended wind bubble, radiation pressure further reduces the efficiency of photoionisation feedback by pushing the ionised gas into a thinner shell, where it recombines faster. This analysis has been further applied to observed feedback structures around massive stars and clusters with an embedded wind bubble in \cite{Pellegrini2009}, \cite{Yeh2012}, \cite{Verdolini2013}, \cite{Rahner2017} and \cite{Pellegrini2020}.

How winds and radiation interact is a timely problem due to the recent inclusion of stellar winds in self-consistent (radiative-)(magneto-)hydrodynamic simulations of star formation on a cloud scale. \cite{Rogers2013} simulated stellar winds from a small cluster of O stars in an inhomogeneous cloud. Meanwhile, \cite{Dale2014} included winds in simulations of stars forming in turbulent clouds with photoionisation. Since then, \cite{Wall2019}, \cite{Wall2020}, \cite{Decataldo2020}, \cite{Geen2020},  and \cite{Grudic2020} have all included stellar winds in star formation simulations, as well as numerous other works studying winds from single sources or in idealised settings \cite[e.g.][]{GallegosGarcia2020} or on larger scales \cite[e.g.][]{Agertz2013,Gatto2017}. To facilitate further development of such studies, more careful work is needed to determine whether simulations properly resolve the interaction between winds, radiation and star-forming clouds.

Simulations of stellar winds are significantly more expensive than simulations with just photoionisation due to the characteristic speeds and temperatures, and the subsequent effect on the simulation timestep. The sound speed in photoionised gas is typically 10 km/s \cite[e.g.][]{Oort1955}, whereas stellar winds have a terminal velocity of up to around 1\% of the speed of light \citep{Groenewegen1989,Lamers1993,Howarth1997,Massey2005}. Given this, careful work is required to explore the parameter space of the influence of winds, and ensure that simulations accurately capture the interaction between each of the processes that occur inside a star-forming cloud. For example, in \cite{Geen2020}, it was noted that the behaviour of the wind bubble had some differences with the Orion nebula described in \cite{Pabst2019}, although there were similarities between the conditions in the two systems. Establishing a firm basis for feedback physics requires a convergence between complex and costly numerical simulations and observations, for which analytic theory is a valuable stepping stone. 


\section{Overview}
\label{overview}

\begin{figure*}
	\centerline{\includegraphics[width=\hsize]{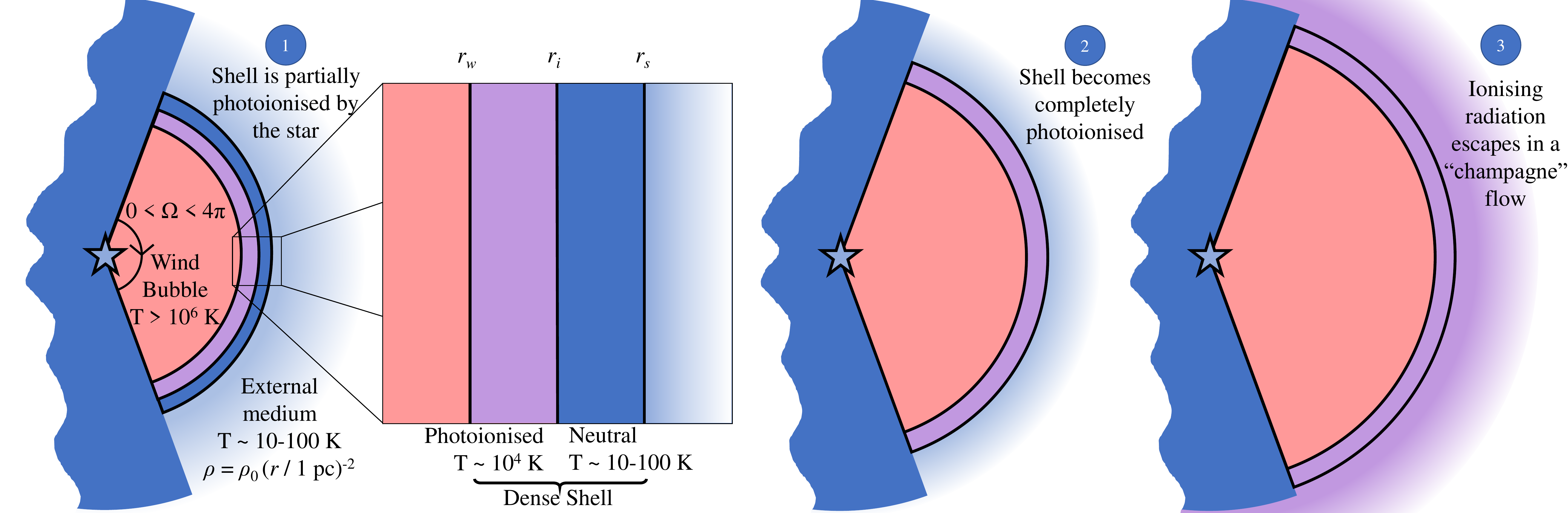}}
    \caption{Diagram showing a schematic of the model used in this paper. 1) A hot ($> 10^6~$K) wind bubble (pink) expands from a stellar source at $r=0$ into a conic section subtended by a solid angle $\Omega$ out to radius $r_w$. If $\Omega < 4 \pi$, a wall of dense gas blocks its expansion in all other directions (blue, left part of the diagram). For the most part we ignore the ```free-streaming'' phase described in \protect\cite{Weaver1977} since it does not affect our result. The wind bubble expands into a neutral external density field with a power law distribution ($\rho = \rho_0 (r / 1 \mathrm{pc})^{-2}$, light blue gradient) and temperature $\sim10-100~$K. A dense shell containing neutral hydrogen forms between $r_i$ and $r_s$ from the material swept up by the wind. The inner part of the shell (mauve) is photoionised by radiation from the star to $\sim 10^4~$K between $r_w$ and $r_i$, while the outer part (blue) between $r_i$ and $r_s$ remains neutral at a similar temperature to the external medium. 2) As the wind bubble expands, more of the shell becomes photoionised until a point where the whole shell is photoionised. 3) The ionising radiation overflows the shell and creates a rapidly-expanding ``champagne'' flow into the external medium. \rev{In this paper we argue that the transition to phase 3 is unlikely in dense molecular cloud conditions until the wind shell is disrupted by some other effect, such as encountering another wind bubble or being broken up by asphericities in the cloud}.}
    \label{fig:diagram}
\end{figure*}

The goal of this paper is to describe the behaviour of feedback structures around young massive stars. Initially, we invoke a wind bubble around the star expanding in a power law density field. Since the medium is dense, it traps ionising radiation initially, which forms a thin photoionised layer inside the dense neutral shell swept up around the wind bubble. This is reasonable because as authors such as \cite{Kuiper2018} find, the initial protostellar outflows around stars are not driven by photoionisation, so it follows that a pre-existing shell trapping ionising radiation can form.

As the dense shell expands, the fraction of the shell that is photoionised becomes larger, until the whole shell is photoionised. After this point, the ionising radiation overtakes the shell around the wind bubble and causes an ``overflow'' of ionising radiation into the surrounding medium, beginning a ``champagne flow'' that rapidly photoionises the unperturbed neutral gas outside.

We aim to describe (i) the early evolution of the wind bubble in a power law density field, (ii) the point at which ionising radiation overflow occurs, and (iii) identify where other effects not included in the main model lead to the requirement for more complex semi-analytic or numerical models.

In Figure \ref{fig:diagram} we show a schematic for the structure of the shell around the wind bubble prior to overflow. The inside of the wind bubble is shock-heated to a high temperature by the energy from winds injected at high velocity by the star. \cite{Weaver1977} describes the internal structure of this wind bubble, which includes a ``free-streaming'' region in the centre of the wind bubble where the material from the wind travels outwards at supersonic speeds before shocking against the gas inside the bubble and creating temperatures > $10^6~$K. We return to the internal structure of the bubble later in the paper where we discuss radiative cooling, but for most of this paper it is only relevant that the bubble is a hot adiabatic volume of gas.

\rev{The interior of the wind bubble is collisionally ionised, and hence the flux of ionising radiation from the star is not noticeably depleted until it encounters the inner edge of the dense shell around the wind bubble at $r_w$ and begins photoionising the gas there.} Between $r_w$ and $r_i$, there is a thin photoionised shell. This gas has a temperature of roughly $10^4~$K, determined by the equilibrium of cooling and heating of the photionised gas. 

In the absence of radiation pressure, this region has a uniform density and is supported by thermal pressure alone. With the inclusion of radiation pressure, there is a pressure differential inside the region due to the larger surface area at larger radii and the presence of dust.

Finally, between $r_i$ and $r_s$, there is a cold, dense neutral shell of material swept up by the wind bubble. At the point of overflow, the quantity of the shell that remains neutral tends to zero, at which point ionising radiation can enter the neutral power law density field outside.

We note that this picture applies for steeper power law density fields as found around young massive stars. In uniform density fields, \rev{\cite{Comeron1997} and} \cite{Silich2013} note that overflow becomes less possible the larger the wind bubble becomes. They argue that as a shell around the wind bubble grows, the \HII region reaches a ``trapping'' point where ionising radiation emitted by the star becomes trapped by the shell around the wind bubble. For much of this paper we thus adopt a fiducial ``singular isothermal sphere'' density field with a power law index of $-2$, which is typical for the conditions around very young massive stars sitting in star-forming cores that have just ended the protostellar phase and is adopted by other authors such as \cite{Shu2002}.

The models in this paper also allow for a wind bubble that is completely constrained by dense gas along certain lines of sight, as in a blister region or a case where outflow occurs only in a small solid angle around the star. In this case we invoke a solid angle subtended by the wind bubble $\Omega$, which tends towards $4 \pi$ as the wind bubble expands in all directions into a spherically symmetric density field.

\subsection{Initial Environment around the star}
\label{overview:environment}

Due to the large parameter space of possible \rev{environments into which feedback from young stars can evolve}, we confine ourselves to a set of physically motivated \rev{quasi-1D models for the} conditions for young massive stellar objects in star-forming cores. \rev{A similar analysis can be applied to other conditions, such as a moving source or a density step function for flows breaking out of a denser cloud environment.}

We define the young massive star as sitting in a cold neutral distribution of gas with a spherically symmetric power law density distribution given by
\begin{equation}
    n(r) = n_0 (r / r_0)^{-\omega}
    \label{eqn:powerlawdensity}
\end{equation}
where $n(r)$ is the hydrogen number density at a radius $r$, $n_0$ is the density at $r_0$ and $\omega$ is the power law index. $\omega=0$ gives a uniform density field, and $\omega=2$ is described as a singular isothermal sphere.

We consider this distribution over an open angle $\Omega$, with a maximum value of $4 \pi$ if the density distribution extends across all lines of sight. Smaller values can be found if much denser gas constrains the bubble along certain lines of sight.  The mass enclosed in radius $r$ in the initial conditions is thus
\begin{equation}
    M(<r) = \int_{0}^{r} \Omega r'^2 n(r') \frac{m_H}{X}.dr' = \frac{\Omega}{3 - \omega} n_0 \, r_0^{\omega} \, r^{3-\omega} \frac{m_H}{X}
    \label{eqn:massinsider}
\end{equation}
where $m_H/X$ is \rev{conversion factor from hydrogen number density to mass density}, and $X$ is the hydrogen mass fraction in the gas. We also define a characteristic mass density $\rho_0 \equiv n_0 m_H / X$.

\rev{\cite{Franco1990} use this distribution above a threshold ``core'' radius of $r_c$, which defines the edge of the protostellar core. For $r < r_c$ they assume that the density is constant, approximating a protostellar core. For the sake of simplicity, we neglect this core radius in further calculations in this paper. We justify this by noting that simulations of protostellar core formation by \cite{Lee2018} find radii where the density profile begins to flatten at radii of around $20~\mathrm{AU}\simeq10^{-4}~\mathrm{pc}$. This is much smaller than the scales studied here (on the order of parsecs).}

\rev{However, the presence of a small core at the point the star is formed remains an important ``initial condition'' in our assumptions due to the fact that in a pure power law density field, ionising radiation would quickly photoionise the whole cloud as in \cite{Franco1990}. We also note that in reality, main sequence stellar feedback is preceded by feedback from protostellar outflows, where photoionisation is a secondary process \citep{Kuiper2018}. Therefore, in a realistic physical context, winds and ionising photons from the star will be emitted into a medium initially preprocessed by a kinetic outflow rather than a pristine neutral gas field.}

\subsection{Stellar Models}
\label{overview:stellar_evolution_models}
 Throughout this work, we use the same stellar evolution tables as \cite{Geen2020}. These are based on \cite{Ekstrom2012} for the stellar parameters, using \cite{Leitherer2014} to give the stellar spectra and a correction from \cite{Vink2011} to convert the escape velocity of the star into a wind terminal velocity. \rev{Relevant physical values for each star are listed in Appendix \ref{appendix:starprops} and Tables \ref{table:app_starprops} and \ref{table:app_starprops_nonrot}.}


\section{Stellar Wind Bubbles}
\label{windmodel}

\rev{In this Section we discuss the evolution of a wind bubble around a young massive star, and the role that environment plays in shaping its evolution.}

\subsection{Revisiting Adiabatic Wind Bubbles}

\cite{Weaver1977} give a solution for the expansion of an adiabatic wind bubble in a uniform medium surrounded by a shell of swept-up interstellar gas. We now re-derive this solution for an adiabatic wind bubble expanding into a power law density distribution, and make observations about how such systems evolve differently to the \cite{Weaver1977} solution.

The dynamical equations governing the wind bubble's expansion are
\begin{equation}
    E_b = \frac{3}{2}\frac{\Omega}{3}r_w ^3 P_w,
    \label{eqn:windenergy}
\end{equation}
\begin{equation}
    \frac{\mathrm{d} }{\mathrm{d} t} \left ( M(<r_w) \frac{\mathrm{d} r_w}{\mathrm{d} t} \right ) = \Omega r_w^2 P_w,
    \label{eqn:windmomentum}
\end{equation}
and
\begin{equation}
    \frac{\mathrm{d E_b} }{\mathrm{d} t} = L_w - \Omega r_w^2 P_w \frac{\mathrm{d r_w} }{\mathrm{d} t},
    \label{eqn:windpressure}
\end{equation}
where $E_b$ is the thermal energy in the shocked gas inside the wind bubble and $P_w$ is the pressure from the wind bubble acting on the inside of the shell. Equation \ref{eqn:windmomentum} describes the conservation of momentum, and Equation \ref{eqn:windpressure} describes the conservation of energy. The first term on the right hand side of this latter equation, $L_w$, is the wind luminosity of the central source, i.e. the mechanical power of the material ejected by the star, assumed constant here for simplicity. The second right hand side term is the work done by the gas pressure $P_w$. The temperature and density of the wind bubble are not immediately important in deriving this solution. We give estimates for these quantities in Section \ref{discussion:cooling}.

\rev{To solve these equations, we require the energy stored inside the wind bubble, which is the fraction of energy emitted by the star not used to do work on the shell around the wind bubble. We derive a quantity for this in Appendix \ref{appendix:analyticenergypartition}, which is given by}
\begin{equation}
    E_b = \left(\frac{5 - \omega}{11 - \omega}\right) L_w t
    \label{eqn:windbubbleenergy}
\end{equation}
\rev{where $t$ is the time since the onset of the wind.}

To produce a tractable solution, we further assume that $r_w(t) \propto t^B$, where $B$ is some power law index. Solving for these equations plus Equation \ref{eqn:massinsider}, we find
\begin{equation}
    r_w(\omega,t) = \left (  A_{w}(\omega,\Omega)  L_w \rho_0^{-1} r_0^{-\omega} t^3  \right )^{1/(5-\omega)}
    \label{eqn:rwt}
\end{equation}
where
\begin{equation}
    A_w(\omega,\Omega) = \frac{4 \pi}{\Omega} \frac{(1-\omega/3)(1-\omega/5)^3}{(1-2\omega/7)(1-\omega/11)} \frac{250}{308 \pi}  
\end{equation}
This becomes the ``Weaver'' solution \citep[equation 21 of ][]{Weaver1977} if $\Omega = 4 \pi$ and $\omega = 0$. \rev{We verify Equations \ref{eqn:windbubbleenergy} and \ref{eqn:rwt} using simple 1D hydrodynamic simulations in Appendix \ref{appendix:numericaltest}, and find good agreement.}

For $\omega=2$ \rev{and $\Omega = 4 \pi$}, Equation \ref{eqn:rwt} becomes
\begin{equation}
    r_{w,2}(t) = 13.9~\mathrm{pc} 
    \left ( \frac{L_w}{10^{36}~\mathrm{erg / s}}  \right )^{1/3}
    \left ( \frac{n_0}{1000~\mathrm{cm}^{-3}}  \right )^{-1/3}
    \left ( \frac{t}{1~\mathrm{Myr}}  \right ),
    \label{eqn:r2}
\end{equation}
which has a constant expansion velocity
\begin{equation}
    v_{w,2}= 13.5~\mathrm{km/s}~\left ( \frac{L_w}{10^{36}~\mathrm{erg / s}}  \right )^{1/3}
    \left ( \frac{n_0}{1000~\mathrm{cm}^{-3}}  \right )^{-1/3}.
    \label{eqn:v2}
\end{equation}

This is a useful result for the regions immediately around young massive stars, since stars typically form in cores where $\omega \simeq 2$ as matter accretes onto the site of protostar formation. \citep[e.g.][]{Lee2018}. At larger radii where this steep profile merges into the cloud background, we expect the solution to become more Weaver-like again, or break up due to clumping and turbulence in the cloud. Where this happens depends on the specific environment around the star.

\subsection{Wind Bubble Expansion Rate}
\label{windmodel:expansionrate}

\rev{In this Section we discuss solutions to Equation \ref{eqn:rwt}.}

\subsubsection{Dependence on Environment}

\begin{figure*}
	\includegraphics[width=\columnwidth]{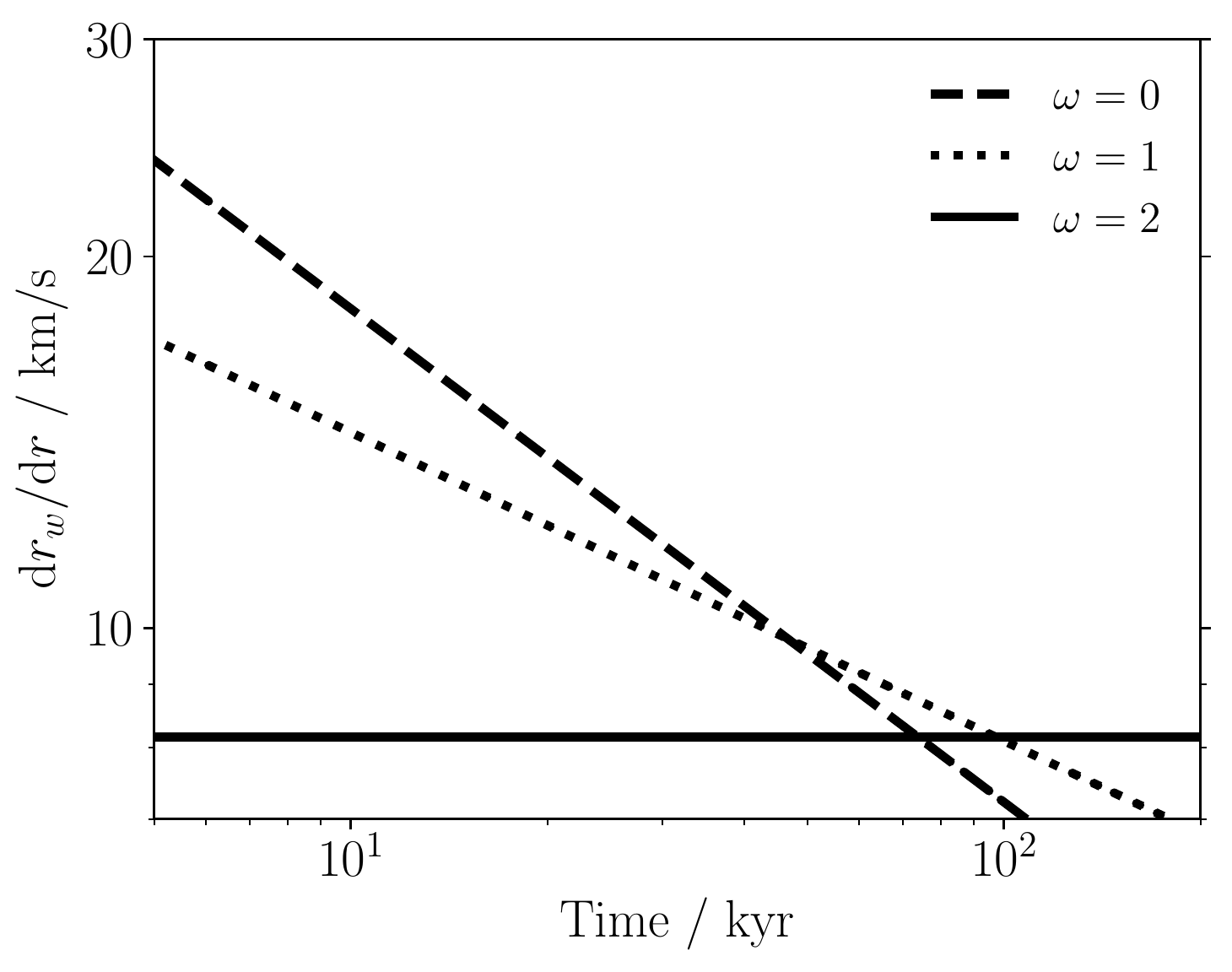}
	\includegraphics[width=\columnwidth]{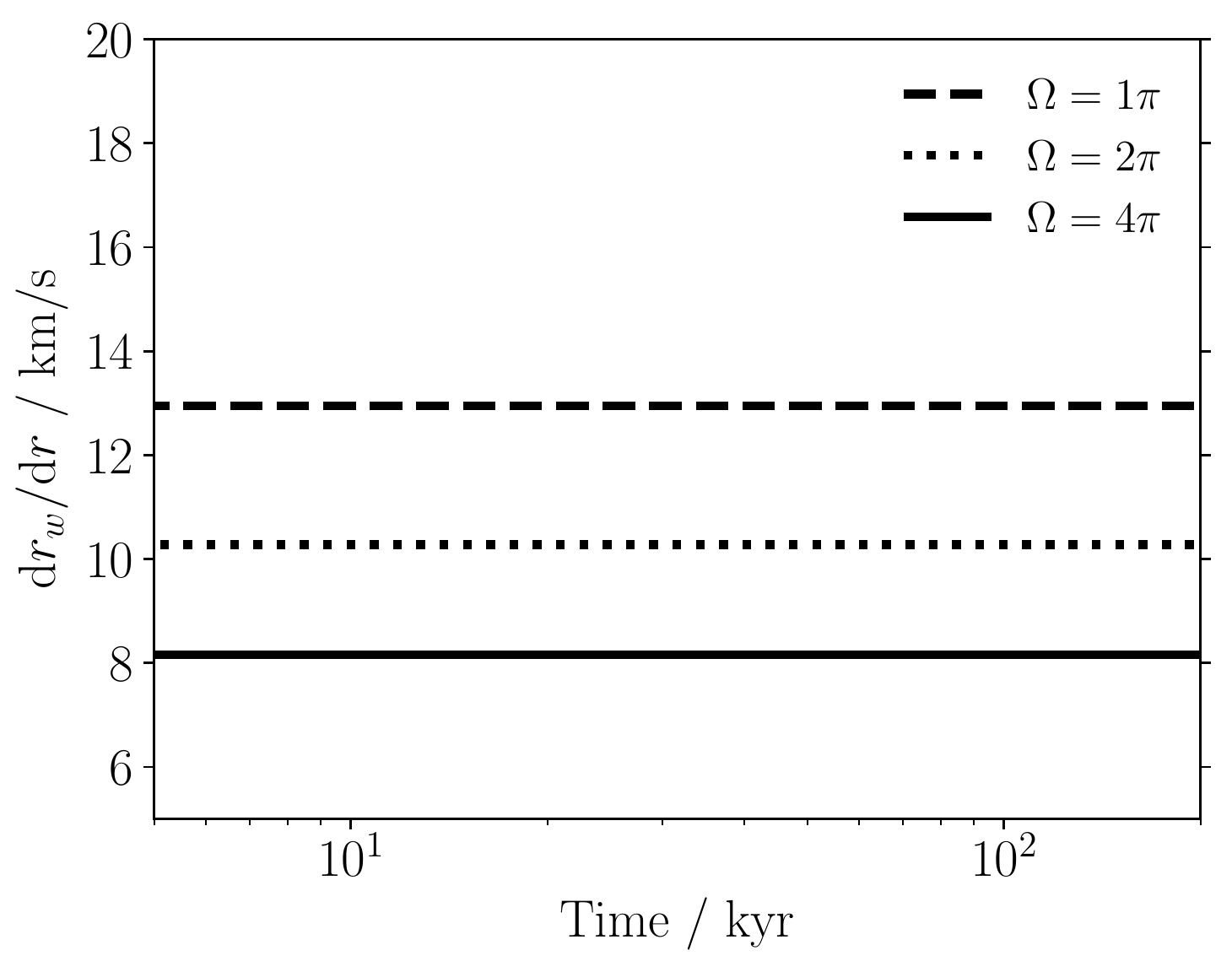}
    \caption{Expansion rate of the wind bubble with time around an example star to demonstrate the role of environment. The left panel shows the density field power law index $\omega$, which describes how steep the density around the star is. The right panel shows the solid angle that the wind bubble expands into, $\Omega$, where $4 \pi$ describes a full sphere with the star at the centre. In the left panel, we fix $\Omega = 4\pi$, and in the right panel we fix $\omega = 2$. For both panels we use a star of mass 35 \Msolar, $n_0=4000~$cm$^{-3}$ and $r_0=1~$pc.}
    \label{fig:drdtforomegaOmega}
\end{figure*}

\rev{We begin by studying the effect of the immediate environment around the star on the behaviour of solutions to Equation \ref{eqn:rwt}. The first parameter is the density field power law index $\omega$, which describes how the density around the star drops with radius. The second is the solid angle subtended by the wind bubble, $\Omega$. We discuss characteristic cloud density $n_0$ below.}

\rev{In Figure \ref{fig:drdtforomegaOmega} we plot the evolution of an example wind bubble around a rotating 35 \Msolar star at Solar metallicity with $\mathrm{log}(L_w/ \mathrm{erg/s}) = 35.9$ (see Table \ref{table:app_starprops}). We use values of $n_0=4000~$cm$^{-3}$ and $r_0=1~$pc, varying $\omega$ and $\Omega$. These values are chosen to approximate the conditions discussed subsequently in the observational comparison in Section \ref{discussion:observations}.}

\rev{The wind velocity is intially higher for shallower power law density fields (lower $\omega$), although this drops over time as the wind bubble expands and the mass of a spherical shell of fixed thickness in the external density field increases. As given in Equation \ref{eqn:v2}, the expansion velocity of a wind bubble is constant where $\omega=2$. We note that the mass of a spherical shell of fixed thickness is constant, and thus both the energy in the bubble and the mass swept up by the wind bubble increase linearly with time. This is a useful result since it removes any time dependence when comparing with observations in young cores where $\omega \simeq 2$, while the radius becomes a simple linear function of the main sequence age of the star.}

\rev{We also plot the expansion velocity of the wind bubble as a function of the solid angle subtended by the wind bubble. A solid angle of $4 \pi$ means that the wind bubble is a sphere centred on the star. However, some wind bubbles expand preferentially in certain directions, being confined by much denser gas in other directions. As the solid angle decreases, the volume of the wind bubble also decreases and hence the pressure inside a bubble of a given radius rises, causing a larger expansion rate.  Equation \ref{eqn:rwt} demonstrates that the shell radius and expansion rate vary with $\Omega^{-1/(5-\omega)}$, so for a power law density field where $\omega = 2$, decreasing $\Omega$ by a factor of 4 (i.e. confining the wind bubble's outflow) increases the wind velocity by around 60\%.}

\subsubsection{Dependence on Stellar Mass, Metallicity and Cloud Density}

\begin{figure}
	\includegraphics[width=\columnwidth]{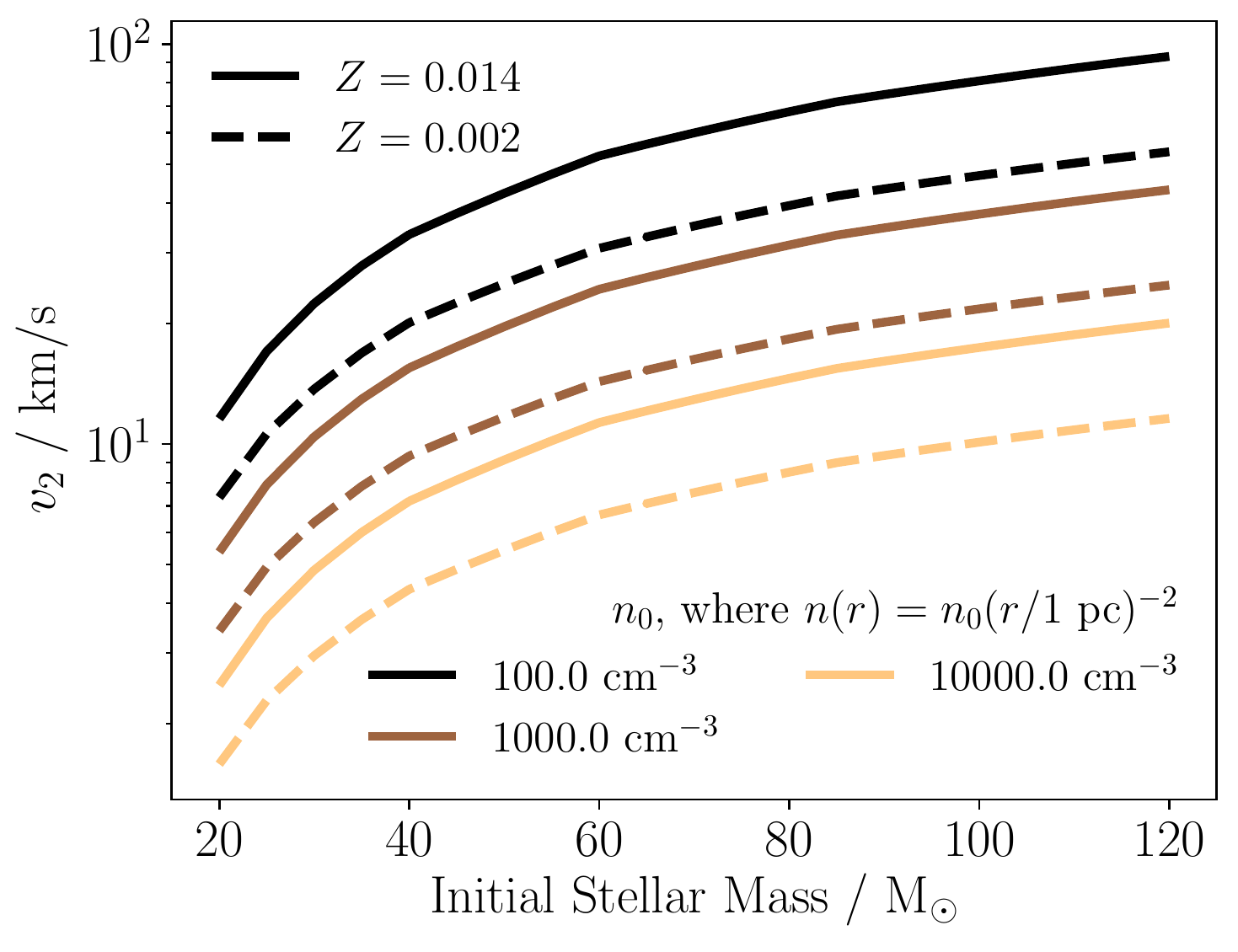}
    \caption{Expansion rate of the wind bubble as a function of stellar mass and characteristic background density $n_0$ at 1 pc where $\omega=2$ and $\Omega = 4 \pi$. Solid lines show results at a Solar metallicity Z=0.014, and dashed lines the results at a sub-Solar metallicity Z=0.002. This expansion rate is constant with time such that $r_{w,2}(t) = v_{w,2} t$. Winds are weaker in the sub-Solar case and so the bubble expands more slowly.}
    \label{fig:drdtforw2Metal}
\end{figure}

\rev{In Figure \ref{fig:drdtforw2Metal} we plot the dependence of the wind bubble expansion rate $v_{w,2}$ in Equation \ref{eqn:v2} (i.e. for a power law density profile where $\omega=2$). Wind luminosity increases with stellar mass and metallicity, giving faster expansion rates. Meanwhile, the expansion rate drops with increasing characteristic density. We include this figure as a quick visual reference for physically relevant conditions.}


\section{Expansion and Overflow of the Photoionised Shell}
\label{ionised}

\rev{The dense shell swept up by the wind bubble interacts with the ionising radiation emitted by the star, and contains an ionised and a neutral component (see Figure \ref{fig:diagram}). In this Section we use the results of Section \ref{windmodel} to calculate the structure and evolution of the photoionised component of the shell, determine at which point the dense shell cannot absorb all of the ionising radiation from the star, and discuss what happens after this point.}

\subsection{Structure of the Ionised Shell}
\label{ionised:structure-of-the-ionised-shell}

\rev{The dense, wind-swept shell is typically dense enough that the recombination time of the gas is short, and hence we assume photoionisation equilibrium, i.e. the photons emitted by the star in the solid angle of the shell are all absorbed by the shell.  We can write the balance of photon emission and absorption as}

\begin{equation}
    \frac{\Omega}{4 \pi} Q_H = \int_{r_w}^{r_i} \Omega r^2 n_i^2(r)\alpha_B.dr 
    \label{eqn:photoionisation_equilibrium}
\end{equation}
where $Q_H$ is the ionising photon emission rate, $n_i(r)$ is the hydrogen number density in the photoionised gas at a radius $r$ and $\alpha_B$ is the \cite{Baker1938} case B recombination rate of the gas for \HII, which is 2 to $3 \times 10^{-13}$ cm$^3$ s$^{-1}$ at solar metallicity. The exact formula used to calculate $\alpha_B$ depends on the temperature of the photoionised gas and is given in Appendix E2 of \cite{Rosdahl2013}.

\rev{Equation \ref{eqn:photoionisation_equilibrium} has an algebraic solution if either (i) the shell is sufficiently thin that $r_i \rightarrow r_w$, or (ii) the density of the photoionised shell is uniform, i.e. $n_i$ is independent of $r$. Case (i) becomes valid where photoionisation is negligible compared to radiation pressure and winds and case (ii) becomes valid where radiation pressure is negligible. A more general solution, however, requires a numerical approach.}

\rev{Numerical solutions for photoionsied shell structures have been used widely in the literature to calculate the dynamical and observable properties of photoionised shells, e.g. \cite{Pellegrini2007}, \cite{Yeh2012} and \cite{Kim2016}. A form of these equations is given in \cite{Draine2011} for dusty \HII regions, which was used by \cite{Martinez-Gonzalez2014} to include a central wind bubble. \cite{Draine2011} gives the conditions for hydrostatic pressure balance, ionising photon balance and dust absorption optical depth respectively at radius $r$ inside the photoionised shell as}
\begin{equation}
n_i \sigma_d \frac{(L_n e^{-\tau} + L_i \phi)}{4 \pi r^2 c} + \alpha_B n_i^2  \frac{\langle h \nu \rangle_i}{c}  - \frac{\mathrm{d} }{\mathrm{d} r} \left (2 n_i(r) k_B T_i   \right ) = 0,
    \label{eqn:draine1}
\end{equation}
\begin{equation}
    \frac{\mathrm{d} \phi }{\mathrm{d} r} = - \frac{1}{Q_H}\alpha_B n_i^2 4 \pi r^2 - n_i \sigma_d \phi,
    \label{eqn:draine2}
\end{equation}
\rev{and}
\begin{equation}
    \frac{\mathrm{d} \tau }{\mathrm{d} r} = n_i \sigma_d,
    \label{eqn:draine3}
\end{equation}
\rev{where $\sigma_d$ is the dust absorption cross section per hydrogen nucleon \citep[we use 10$^{-21}~$cm$^{-2}$ $Z / Z_{solar}$ as in][where $Z$ is the gas metallicity and $Z_{solar}=0.014$]{Draine2011}, $L_n$ and $L_i$ are the luminosities of the star for non-hydrogen-ionising and hydrogen-ionising photon energies respectively, $\phi$ is the fraction of ionising photons reaching radius $r$, $c$ is the speed of light, $\langle h \nu \rangle_i$ is the average energy of ionising photons (where $h \nu$ > 13.6 eV), $T_i$ is the temperature of the photoionised gas (where $2 k_B T_i = c_i^2 m_H / X$) $\simeq 10^4~$K, and $\tau$ is the dust absorption optical depth. }

\rev{We solve these equations beginning at $r_w$, where the density $n_i$ is set by the pressure balance with the wind bubble, }
\begin{equation}
    P_w = P_i \equiv \frac{m_H}{X} n_i c_i^2
    \label{eqn:PwPibalance}
\end{equation}
\rev{\citep[e.g.][]{Dyson1980}, where $P_w$ is given in Equation \ref{eqn:windpressure}. We integrate for increasing $r$ until $\phi$ reaches zero, i.e. all the ionising photons have been absorbed, which we define as the edge of the photoionised shell $r_i$. }

For each value of $r_w$ we use, we calculate the mass of the photoionised shell 
\begin{equation}
    M_i = \int_{r_w}^{r_i} \Omega r^2 n_i(r) \frac{m_H}{X}.dr 
    \label{eqn:photoionised_mass},
\end{equation}
and compare it to the total mass of gas swept up by the \HII region out to $r_i$, $M(<r_i)$ using Equation \ref{eqn:massinsider}. If the value found for Equation \ref{eqn:photoionised_mass} exceeds $M(<r_i)$, the solution breaks down because there is no more mass in the neutral shell to absorb the ionising photons. We refer to this stage of the solution as ``overflowed'', since ionising photons are able to overflow the shell. We discuss the consequences of this later in this Section.

\subsection{Numerical Solutions for Shell Structure and Overflow}
\label{ionised:results}

\begin{figure*}
	\includegraphics[width=\columnwidth]{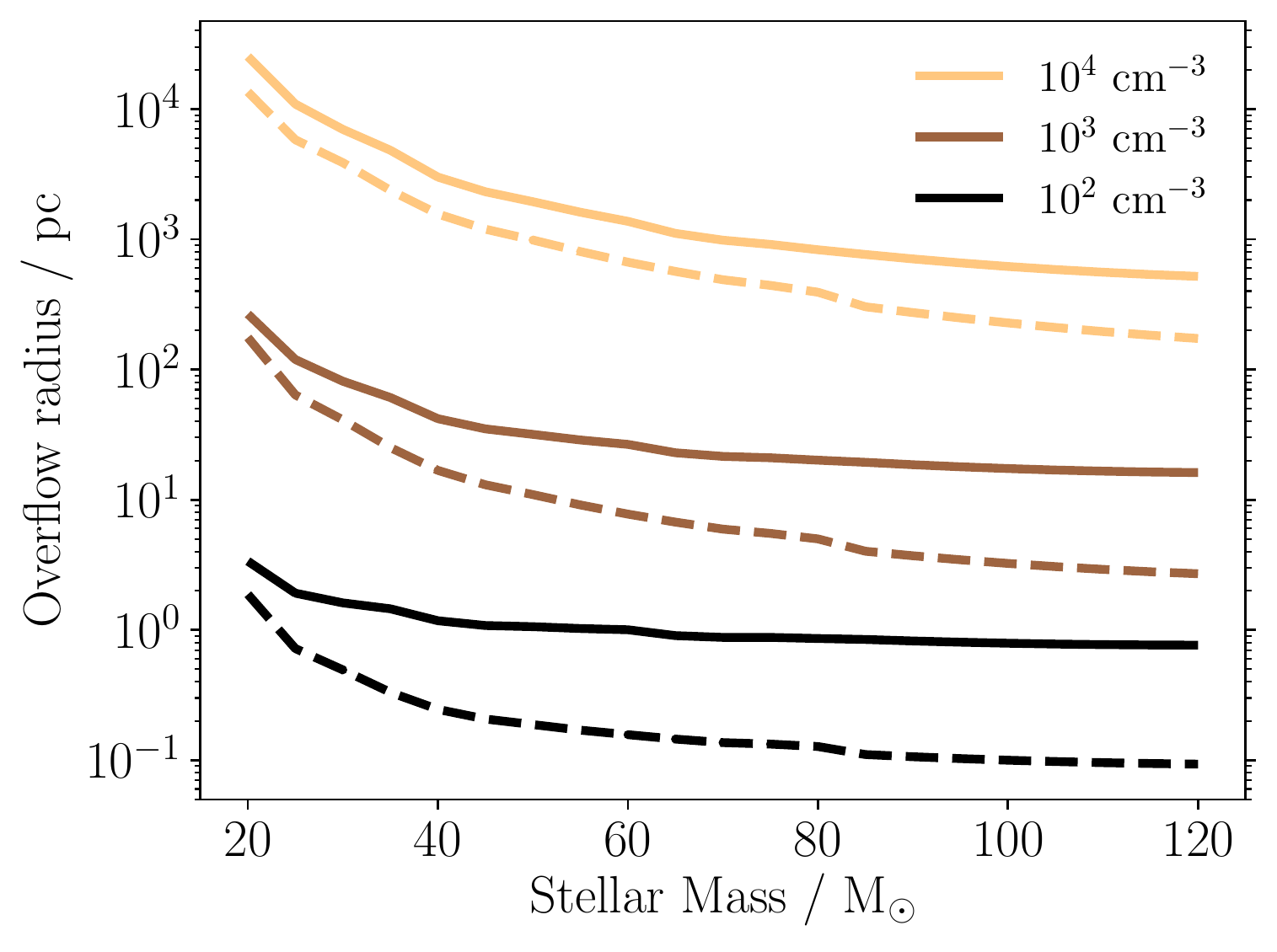}
	\includegraphics[width=\columnwidth]{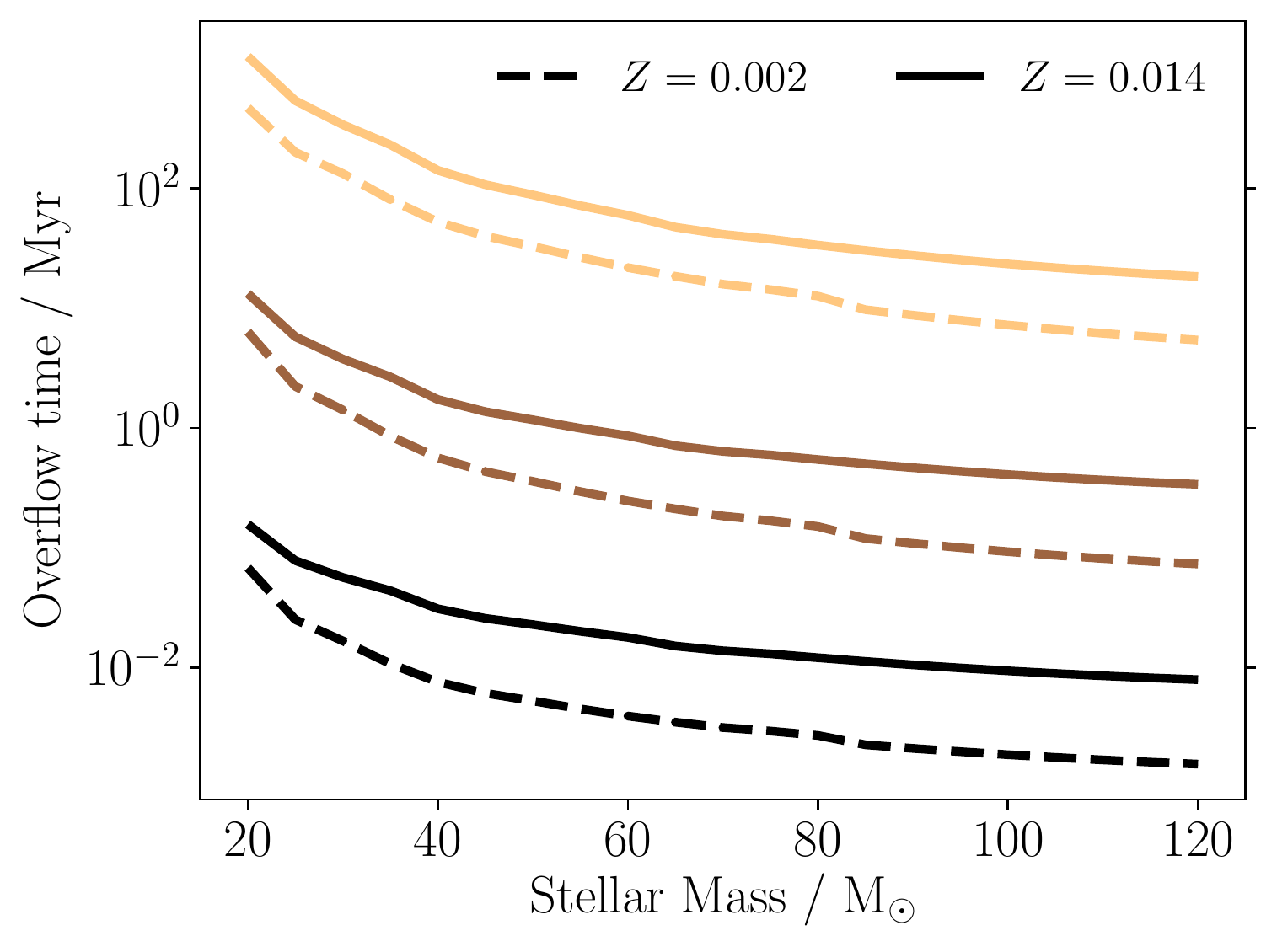}
    \caption{\rev{Point at which ionising photon overflow occurs as a function of initial stellar mass, with varying metallicity} and characteristic background hydrogen number density $n_0$, where $n(r) = n_0 (r~/~1~\mathrm{pc})^{-2}$. The left plot shows the outer radius of the ionised shell $r_i$ at which overflow occurs, the right plot shows the time of overflow. The denser the environment, the denser the shell around the wind bubble and hence the longer it takes for ionising photons to overflow the shell. Due to weaker winds, overflow occurs sooner and at smaller radii at lower metallicity. \rev{Solid lines show results at Solar metallicity ($Z=0.014$) and dashed lines show results at sub-Solar metallicity ($Z=0.002$). Line colour shows characteristic cloud density $n_0$.}}
    \label{fig:overflowradii_times}
\end{figure*}

\begin{figure*}
	\includegraphics[width=\columnwidth]{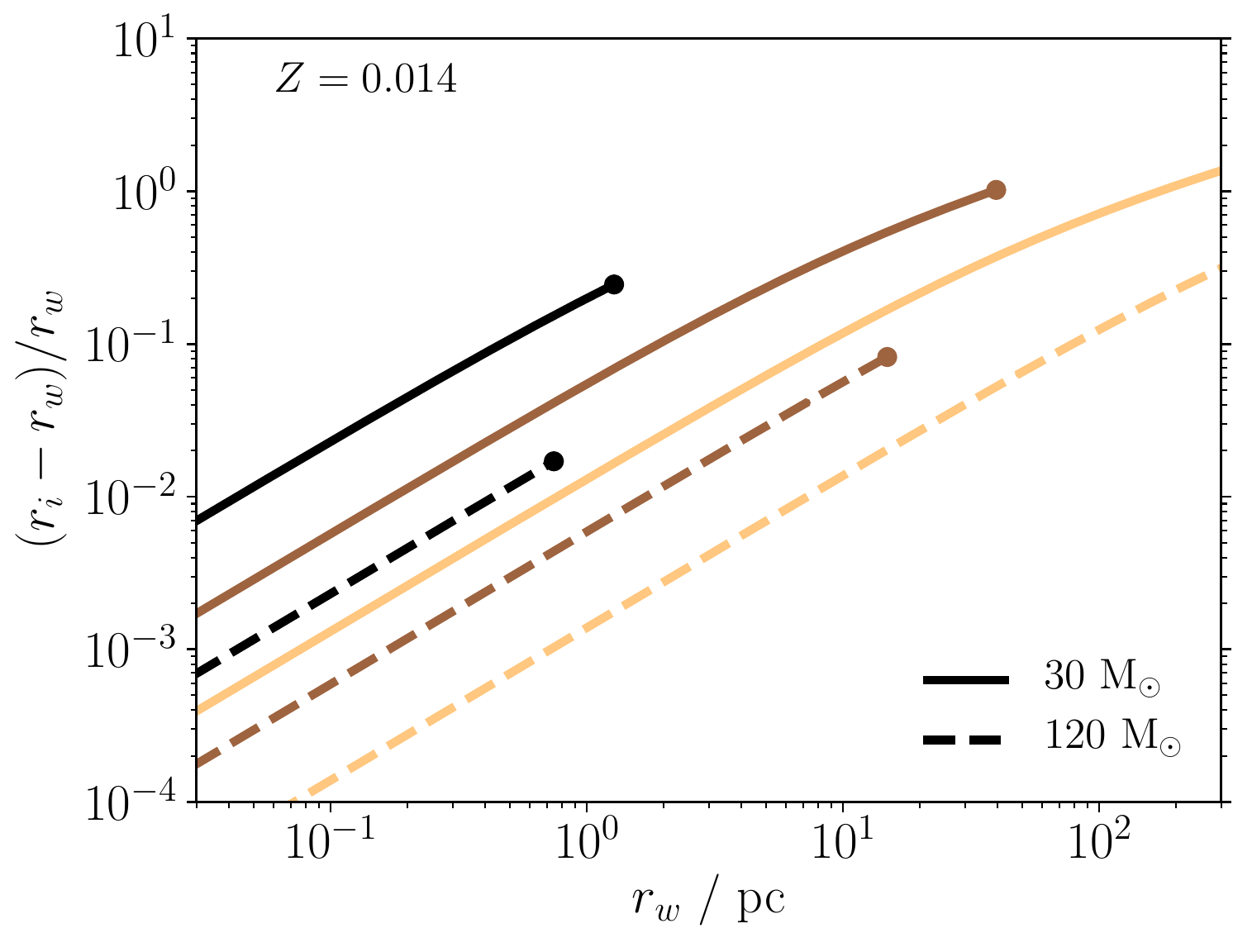}
	\includegraphics[width=\columnwidth]{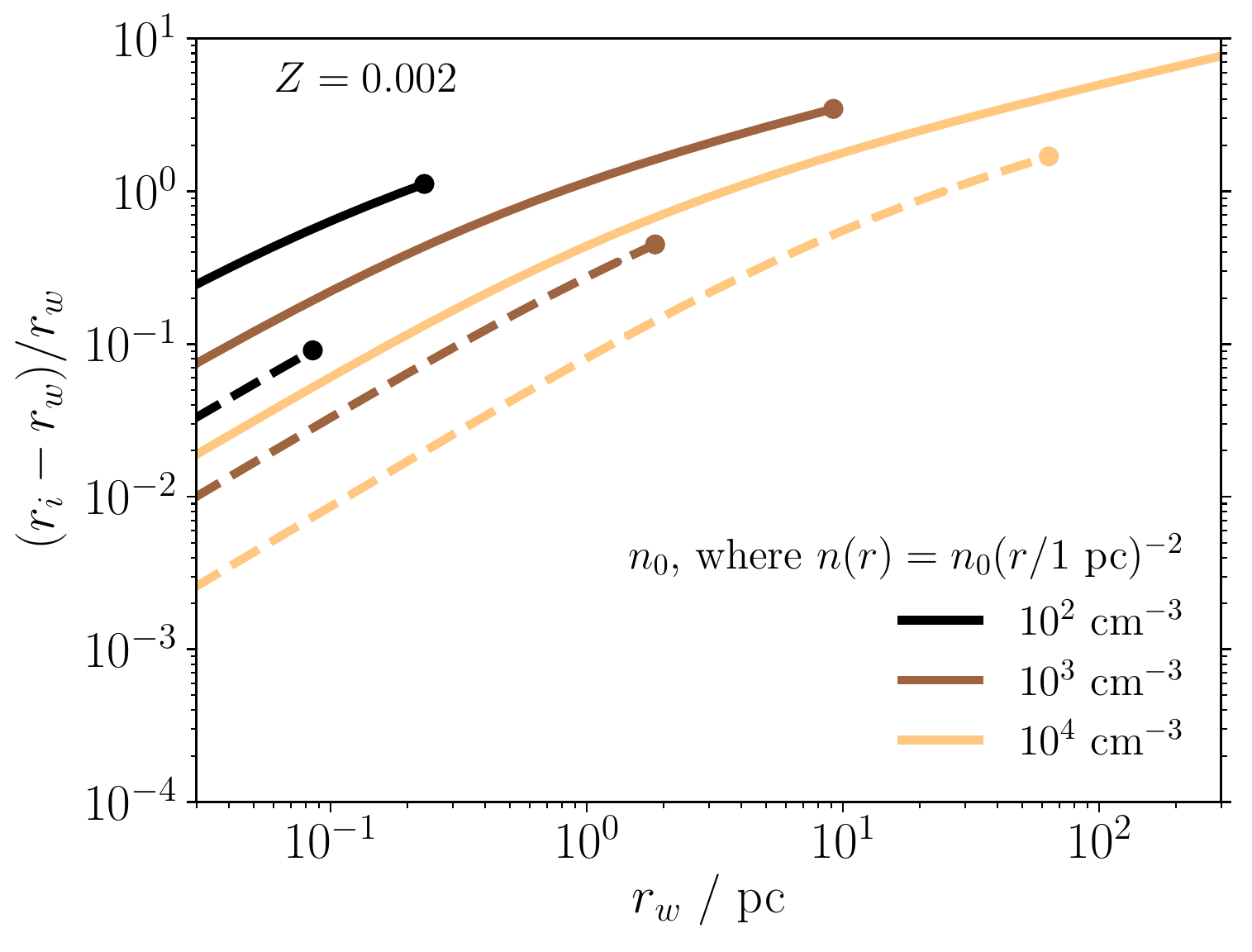}
    \caption{\rev{Relative thickness of the photoionised shell $r_i - r_w$ against the wind bubble radius $r_w$, plotted as a function of $r_w$ with varying initial stellar mass and characteristic background hydrogen number density $n_0$, where $n(r) = n_0 (r~/~1~\mathrm{pc})^{-2}$. The left plot shows results at $Z=0.014$, and the right plot shows results at $Z=0.002$. Dashed lines show results for a 120 \Msolar star, solid lines show results for a 30 \Msolar star. Line colour shows characteristic cloud density $n_0$. The relative thickness of the ionised shell is typically small until near the point of overflow, which is shown as a circle at the end of each line. We omit results after 250 pc since this is larger than most molecular clouds in our Galaxy and nearby galaxies. We also note that substructure in complex star-forming environments is likely to cause the power law density distribution to break down at smaller radii than this. In this scenario, photon breakout will be caused by 3D effects before the \HII region reaches the overflow point.}}
    \label{fig:ionisedshellthickness}
\end{figure*}

\rev{In this section we plot a set of solutions to the equations in Section \ref{ionised:structure-of-the-ionised-shell}. In order to provide a tractable subset of the parameter space of possible inputs, we focus on cases where $\omega = 2$ for a set of physically-motivated conditions. As in Section \ref{windmodel:expansionrate}, we use values from Table \ref{table:app_starprops} for the stellar properties. For these solutions, we require $Q_H$, $L_n$, $L_{i} \equiv Q_H \langle h \nu \rangle_i$ , $T_i$ and $L_w$. The latter quantity is used to calculate $r_w$, which is the inner boundary of the photoionised shell. We use fiducial densities $n_0 = 10^2, 10^3~$and~$10^4~$cm$^{-3}$, where $n(r) = n_0(r / 1~$pc$)^2$ for the purposes of this analysis.}

\rev{Figure \ref{fig:overflowradii_times} shows the time and ionised shell radius at which the ionising radiation overflows the shell (i.e. there is not enough mass in the shell to absorb all the ionising radiation). There is a small dependence on metallicity, and a much larger dependence on cloud density. Note that the lifetime of the stars described here is on the order of 10 Myr, or less in the case of very massive stars. We thus expect the star to reach the end of its main sequence before overflow conditions are reached in very dense clouds. By contrast, for more diffuse clouds, overflow may occur when the star is younger. We discuss other scenarios outside the scope of the 1D model in Section$~$\ref{discussion}. Nonetheless, this model suggests that, provided certain plausible initial conditions are achieved, there is a range of physically relevant conditions for which the ionising radiation from massive stars is trapped for some or all of a star's life, preventing a champagne flow and reducing the influence of ionising radiation in molecular clouds.}

\rev{We plot the relative thickness of the ionised shell compared to the radius of the wind bubble, $(r_i - r_w) / r_w$ in Figure \ref{fig:ionisedshellthickness}. This is initially small compared to the overall size of the wind bubble, growing towards the point of overflow to be similar in radius, or even larger for sub-Solar metallicities. The relative thickness is smaller for more massive stars due to the stronger winds. The relative expansion velocity of the ionised shell compared to the wind velocity $(\dot{r}_i - v_{w,2}) / v_{w,2}$ follows a very similar relation, having values between 1.5 and 2 times $(r_i - r_w) / r_w$. Therefore, the expansion rate of the edge of the ionised shell, where C$^{+}$ emission is located, begins at around $v_{w,2}$, but can be up to $2 v_{w,2}$ at the point of overflow.}

\rev{The effect of radiation pressure on the structure of the ionised shell is typically small compared to the influence of winds. We can measure this by measuring the density gradient in a photoionised region around a star \citep{Draine2011,Martinez-Gonzalez2014}. If radiation pressure is non-negligible, the density should be constant with radius. This is because the pressure balance in the photoionised gas can be written as}
\begin{equation}
    P_i(r) = \rho_i(r) c_i^2 + P_{rad}(r),
\end{equation}
\rev{where $P_i$ is the total pressure at $r$ in the photoionised gas and $P_{rad}$ is the contribution from radiation pressure at $r$. For a hydrostatic shell, $P_i$ is constant with $r$, as is $c_i$ due to the cooling equilibrium in photoionised gas. $P_{rad}$ drops as the flux of radiation decreases over the radius of the shell due to geometric dilution and absorption by atoms and dust. Therefore, in order to maintain a constant $P_i$, $\rho_i$ must increase with $r$.}

\rev{\cite{Draine2011} measures the density contrast in a photoionised region by comparing the root-mean-square (RMS) density, the mean density and the maximum density, found at the outer edge of the region at $r_i$. The ratio between the RMS density and the mean density is negligible in all cases in our model. The maximum density is initially 1-5\% higher than the RMS density at Solar metallicities, dropping towards smaller values for larger wind bubbles where the radiation pressure is spread across a larger surface area. For $Z=0.002$, this difference is initially 2-13\%, again dropping towards negligible values at larger radii. Radiation pressure can thus have a second-order effect on the structure of compact \HII regions, but this importance drops at larger \HII region radii towards the point of overflow.}

\subsection{Analytic Description}

\rev{In order to interpret the results in Section \ref{ionised:results}, we produce an approximate and illustrative analytic solution for the expansion of the photoionised shell that allows us to more clearly explain the behaviour of the numerical solutions to Equations \ref{eqn:draine1} to \ref{eqn:draine3} presented in Section \ref{ionised:results}.}

\subsubsection{Ionised Shell Structure}
\label{ionised:structure-of-the-ionised-shell-analytic}

\rev{Our first assumption is that radiation pressure causes only a small change in $n_i$, and thus we assume a constant $n_i$, giving a solution to Equation \ref{eqn:photoionisation_equilibrium} that reads}
\begin{equation}
    Q_H = \frac{4 \pi}{3} n_i^2 (r_i^3 - r_w^3) \alpha_B,
    \label{eqn:photoionisation_equilibrium_uniform}
\end{equation}
\rev{noting the dependence on $4 \pi$ and not $\Omega$ since the radiation travels in all directions, not just the solid angle subtended by the wind bubble.}

\rev{Using Equation \ref{eqn:PwPibalance} and substituting for $P_w$ using Equations \ref{eqn:windenergy} and \ref{eqn:windbubbleenergy}, we can write the number density of the photoionised gas as}
\begin{equation}
    n_i = \frac{2}{\Omega r_w^3} \left( \frac{5-\omega}{11-\omega}\right) \frac{X}{m_H}\frac{1}{c_i^2}L_w t.
\end{equation}

\rev{Substituting for $n_i$ in Equation \ref{eqn:photoionisation_equilibrium_uniform}, we can write}
\begin{equation}
    r_i^3 = r_w^3 + \frac{3}{4 \pi} \frac{Q_H}{\alpha_B} \left( \frac{\Omega}{2} \left( \frac{11-\omega}{5-\omega}\right)\frac{m_H}{X} \frac{c_i^2}{L_w t} \right)^2 r_w^6 
\end{equation}

\rev{Substituting for $t$ in Equation \ref{eqn:rwt}, we can write this as }
\begin{equation}
    r_i = r_w\left(  
    1 + 3 B_w~
    r_w^{(2\omega - 1)/3}
    \right)^{\frac{1}{3}}
    \label{eqn:rivsrw}
\end{equation}
\rev{where}
\begin{equation}
    B_w = \pi \left(\frac{\Omega}{4 \pi}\right)^2
    \left( \frac{11-\omega}{5-\omega} \right)^2
    \left(\frac{m_H}{X}\right)^2 
    \frac{Q_H}{\alpha_B} c_i^4 
    \left(\frac{A_w(\omega,\Omega)}{L_w^2\rho_0 r_0^{\omega}}\right)^{2/3}
\end{equation}

\rev{In the thin shell limit (where $r_i \rightarrow r_w$), this equation can be written using the binomial approximation as }
\begin{equation}
    \Delta r \equiv r_i - r_w \simeq 
    B_w~
    r_w^{(2\omega - 1)/3}
\end{equation}

\rev{Some general comments may be made about the form of this equation. Firstly, the thickness of the ionised shell increases with $r_w$ for $\omega > 1/2$, and decreases for $\omega < 1/2$. Secondly, the thickness increases with photoionisation-related quantities $Q_H$ and $c_i$ and decreases with the wind luminosity $L_w$ and characteristic background density $\rho_0$.}

\rev{It is also possible to calculate the thickness of the neutral shell from the remaining swept-up mass that is not photoionised. This also requires a cooling function to calculate its density from the hydrostatic pressure in the shell, which in turn allows an estimate of the thickness, as well as including the effects of FUV radiation pressure on dust in the shell. However, we do not discuss this here since it does not affect the results of this work.}

\subsubsection{Condition for Overflow}
\label{ionised:condition-for-overflow}

\rev{In order to calculate the overflow radius, we make a further simplifying assumption that the mass swept up by the shell $M(<r_i) \simeq M(<r_w)$. If we do not apply this assumption, a cubic equation in $r_w$ can be found by using Equation \ref{eqn:rivsrw} that gives a similar result, although this becomes more difficult to interpret.}

\rev{Since the density inside the wind bubble is typically very low, the mass in the dense shell around the wind bubble is assumed to be equal to the swept-up mass of unshocked gas at $t=0$ in the background medium, i.e. $M(<r_w)$ using Equation \ref{eqn:massinsider}. In the limit where this mass equals the mass of the photoionised shell, all of the shell is photoionised. Beyond this point, the shell cannot absorb all of the ionising photons, which begin leaking into the surrounding neutral gas. We can write a condition for such photon ``overflow'' as}
\begin{equation}
    \frac{\Omega}{3 - \omega} n_0 r_0^{\omega} r_w^{3-\omega} \frac{m_H}{X}
    \le
    \frac{\Omega}{3} \left(r_i^3 - r_w^3 \right) n_i \frac{m_H}{X}
\end{equation}

\rev{Making the same substitutions for $n_i$ and $t$ as in Section \ref{ionised:structure-of-the-ionised-shell}, we can produce an expression for the wind bubble radius at which overflow is possible, where}
\begin{equation}
    r_{w}^{\frac{5-4 \omega}{3}} \le 
    \frac{\Omega}{4 \pi} 
    \left( \frac{3-\omega}{2} \right)
    \left( \frac{11-\omega}{5-\omega} \right)
    \frac{Q_H c_i^2}{\alpha_B}
    \left(\frac{m_H}{X} \frac{A_w^{1/2}(\omega,\Omega)}{L_w n_0^2 r_0^{2\omega}} \right)^{2/3}
    \label{eqn:overflowcondition}
\end{equation}

\rev{We define an overflow radius $r_{o}$ equal to the radius $r_w$ at the limit where overflow is just possible, i.e. the mass of the photoionised shell is exactly equal to the mass of the swept-up material.}

\rev{There are again some major points to note concerning the behaviour of this equation. The right hand side increases for larger values of $Q_H$ and $c_i$, and decreases for larger values of $L_w$ and $n_0$. However, the left hand side behaves differently depending on the value of $\omega$. For $\omega < 5/4$, as in an initially uniform density field where $\omega=0$, overflow becomes \textit{less} likely with increasing radius, assuming a dense shell can accumulate at $r_w$ rather than being dispersed by hydrodynamic mixing. This mode is described in \cite{Comeron1997} and \cite{Silich2013}}, who argue for a ``trapping'' radius, where ionising radiation initially forms an extended photoionised region, but over time a wind shell grows \rev{in the warm photoionised medium at $r_w$ until its density and mass are sufficient to trap the radiation}.

\rev{However, for a power law density field above this value, such as where $\omega = 2$, the left hand side decreases with radius, and so overflow becomes \textit{more} likely as the wind bubble grows.}

\rev{This is important because for a wind bubble around a young massive star expanding into a steep power law density field there exists a family of solutions for which the ionising radiation cannot escape until some overflow radius. This requires that the initial conditions are such that an instant champagne flow as described by \cite{Franco1990} is prevented, e.g. by a small shallow core in the density profile around the star or an initial shell set up by protostellar outflows that can trap ionising radiation during the main sequence stage of the star's evolution. If these conditions are met, this offers a plausible explanation for why we observe some young \HII regions that do not exhibit photoionised champagne flows, such as M42 in Orion \citep{Pabst2019}, and rather appear to be filled with hot, wind-shocked gas \citep{Guedel2008} and are surrounded by neutral shells \citep{Pabst2020}.}

\rev{In the following Section, we discuss perturbations to the model from other physical effects, and cases where photon escape may be possible before the overflow radius is reached.}

\subsection{Post-Overflow Conditions}

\rev{Reaching the overflow radius does not guarantee an immediate and strong champagne flow. Firstly, just after the point of overflow, while the neutral shell and any observational tracers from it disappear, only a small number of ionising photons will leak into the surrounding medium, causing only a weak decoupling of the thermally-expanding \HII region and ionisation front. Initially, the shell does not respond to the change in conditions outside. However, as the \textit{rarefaction} wave that moves backwards from $r_i$ approaches $r_w$, pressure gradients will cause the shell around the wind bubble to begin to be dispersed into the surrounding medium. Under these conditions, the recombination rate in the shell drops and more ionising photons can leak, causing the shell to disintegrate and a strong champagne flow to begin. This will likely be enhanced if discontinuities in the cloud density are encountered, e.g. as the shell reaches a sharp step in density between the molecular cloud and the diffuse interstellar medium around it, as in \cite{TenorioTagle1979}, or if clumping in the cloud breaks up the wind-driven shell. Under these conditions, the wind bubble will likely rapidly break out of the cloud in an unstable plume as found in \cite{Comeron1997} and \cite{Geen2020}.}

\rev{For very massive stars in more diffuse molecular clouds, the wind bubble can expand much faster than the speed of sound in the photoionised gas, so under these conditions the wind bubble can expand more rapidly than the gas can expand thermally after it is photoionised, preventing a strong champagne flow. The precise behaviour of the \HII region after it reaches the overflow radius is mostly beyond the scope of this paper, since we primarily wish to explain why certain \HII regions appear to retain neutral shells that suggest they precede the emergence of a champagne flow. We thus leave a more detailed description of the post-overflow dynamics for future work.}

\section{Discussion}
\label{discussion}

In this Section we discuss effects neglected so far in our model, and the impact they may have on the solutions.

\subsection{Magnetic Fields}

In this paper so far we have neglected magnetic fields. We now discuss their likely effect on the evolution of the shell around the wind bubble.

We can modify Equation \ref{eqn:PwPibalance} to include a term to describe the magnetic pressure in the shell. We then get
\begin{equation}
    P_w = \frac{m_H}{X} n_i(r_w) c_i^2 + \frac{B^2}{8 \pi},
    \label{eqn:innerpressure_Bfield}
\end{equation}
where the first term of the right hand side is the thermal pressure of the inner edge of the shell $P_{therm}$ and the second is the magnetic pressure $P_{mag}$.

To get an estimate of the likely magnetic field in the shell, we use an approximation to \cite{Crutcher2012} and assume a critical B-field, where
\begin{equation}
    \left ( \frac{B}{10~\mu \mathrm{G}} \right )^2 = \frac{n_i(r_w)}{300~\mathrm{cm}^{-3}}
\end{equation}
in the regime studied here. The magnetic field saturates at $10~\mu \mathrm{G}$ below 300 cm$^{-3}$. Equation \ref{eqn:innerpressure_Bfield} then becomes:
\begin{equation}
    P_w = \frac{m_H}{X} n_i(r_w) c_i^2 + (10~\mu \mathrm{G})^2 \frac{n_i(r_w)}{300~\mathrm{cm}^{-3}}.
    \label{eqn:innerpressure_Bfield_withdensity}
\end{equation}

The plasma beta ($\beta \equiv P_{therm} / P_{mag}$) of the ionised shell assuming $c_i \simeq $10 km/s is 169. The pressure from magnetic fields is thus under 1\% of the thermal energy, and so we justify neglecting it for the rest of the calculations in the paper. The importance of magnetic fields is likely instead to be to prevent the fragmentation of the shell and other gaseous structures, as found in \cite{Hennebelle2013} and \cite{Geen2015b}. Models taking into account geometry effects on the magnetic field may also lead to a non-negligible role in shaping the photoionised shell, as discussed in \cite{Pellegrini2007}.

\subsection{Gravity}
\label{discussion:gravity}

\begin{figure}
	\includegraphics[width=\columnwidth]{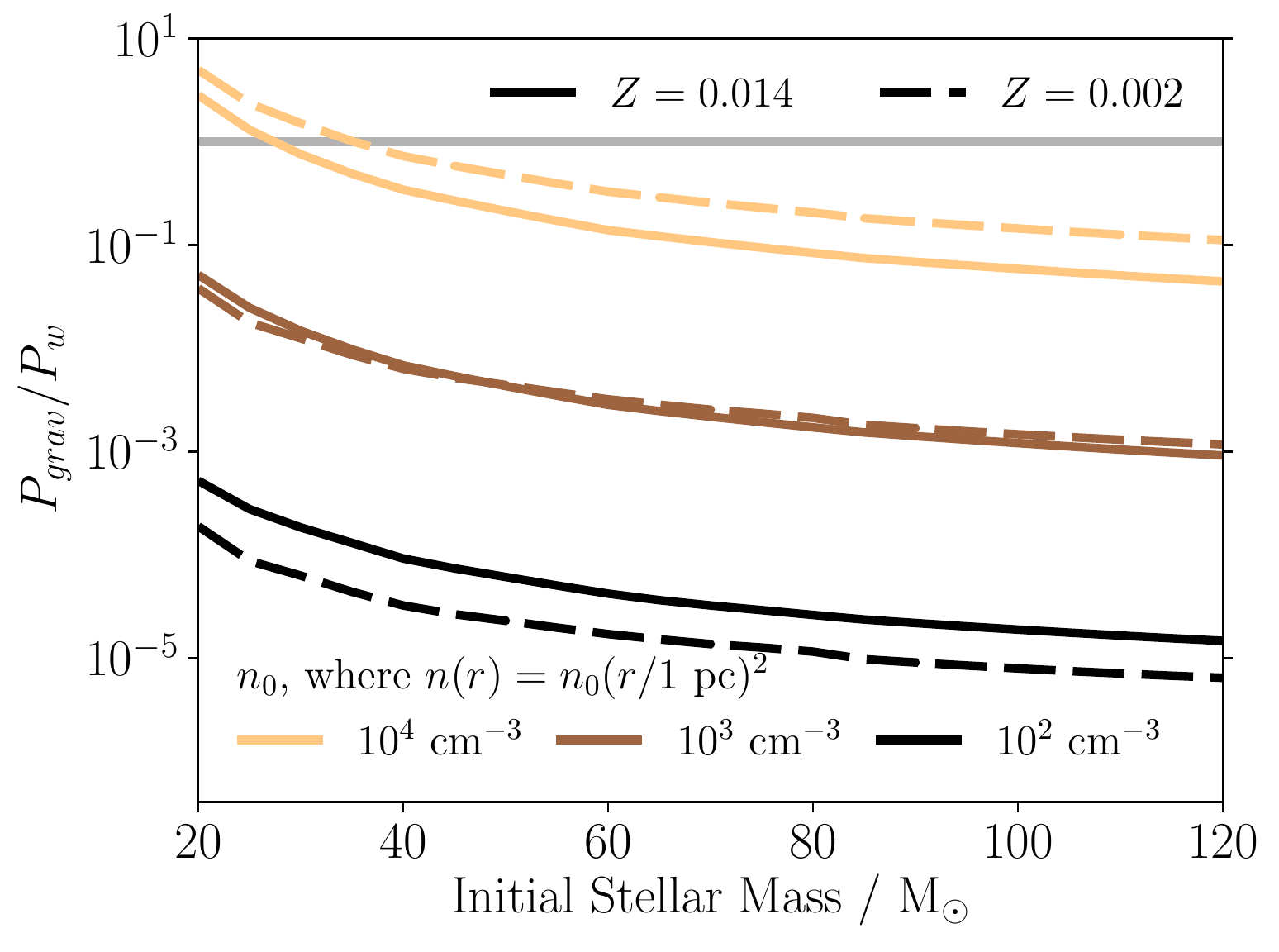}
    \caption{Ratio of pressure from gravity on the shell compared to the pressure from the wind bubble as a function of stellar mass and characteristic background density $n_0$ at 1 pc where $\omega=2$. Gravity is more important in denser environments, but typically does not overcome the pressure inside the wind bubble for this range of densities. \rev{A horizontal grey line shows the point where $P_{grav} = P_{w}$}.}
    \label{fig:PgravvsPbubble}
\end{figure}

We have until now ignored gravity in our model. The pressure from gravity on a spherical shell of inside \rev{$r_i$} is given by 
\begin{equation}
    P_{grav} = \frac{G M(<r_w)^2}{4 \pi r_i^4}
\end{equation}

We calculate $P_{grav} / P_w$ at the point of overflow for the same conditions as Figure \ref{fig:overflowradii_times} and plot it for various stellar masses and densitites in Figure \ref{fig:PgravvsPbubble}. Gravity has only a small effect at low densities, high metallicities and large stellar masses. \rev{Only for the highest densities and lowest stellar masses does gravity become comparable to the pressure from winds. We therefore justify ignoring it for this work, although in certain regimes, particularly if wind cooling becomes efficient and hence $P_w$ drops, it may act to compress the shell or even stall the expansion of the \HII region.}

\rev{For regimes such as older \HII regions around massive clusters}, gravity has been found to be more important where the pressure inside the bubble has begun to drop due to expansion and radiative cooling \citep[e.g.][]{Rahner2017}.

\subsection{Radiative Cooling}
\label{discussion:cooling}

\begin{figure*}
	\includegraphics[width=\columnwidth]{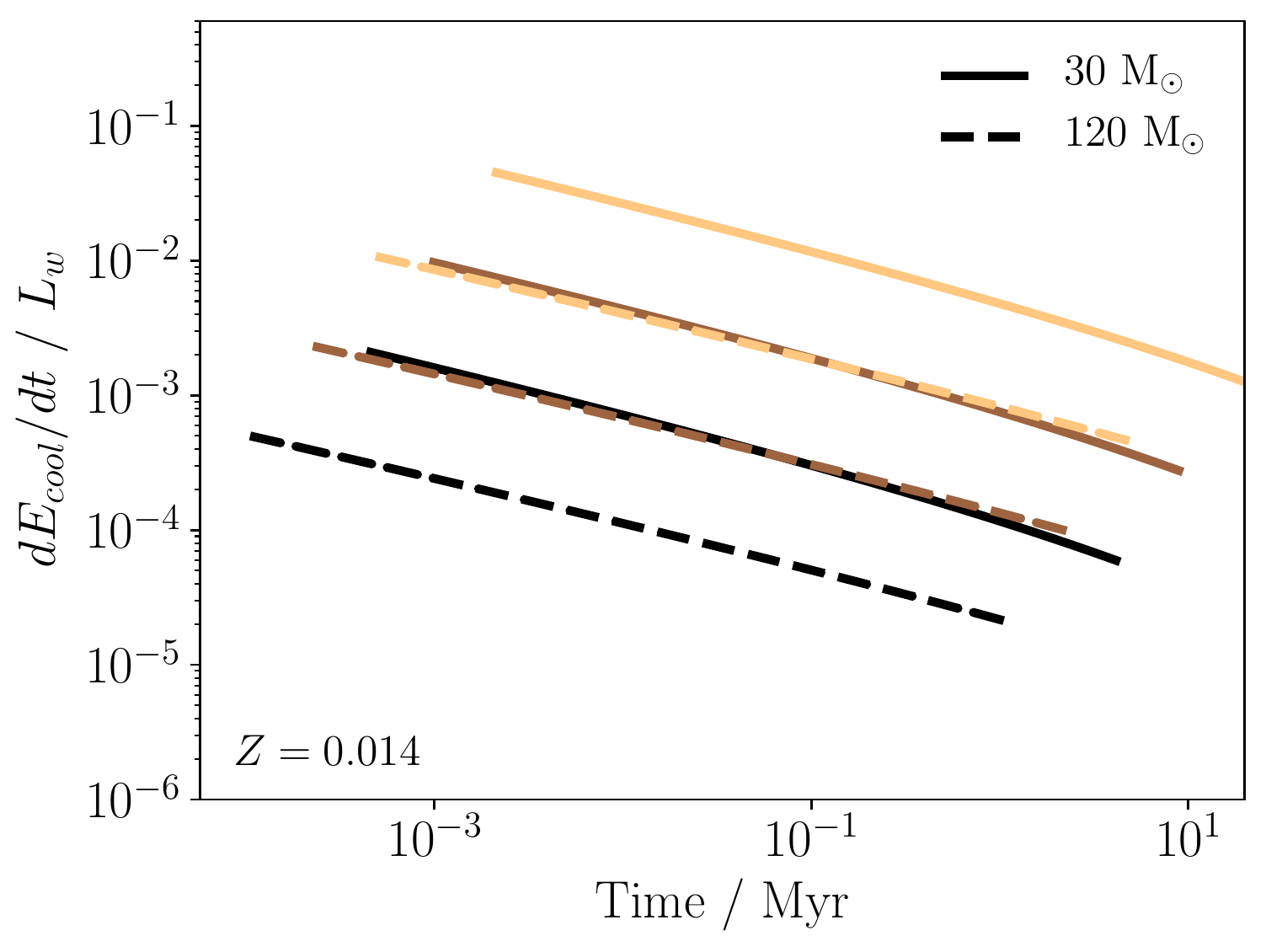}
	\includegraphics[width=\columnwidth]{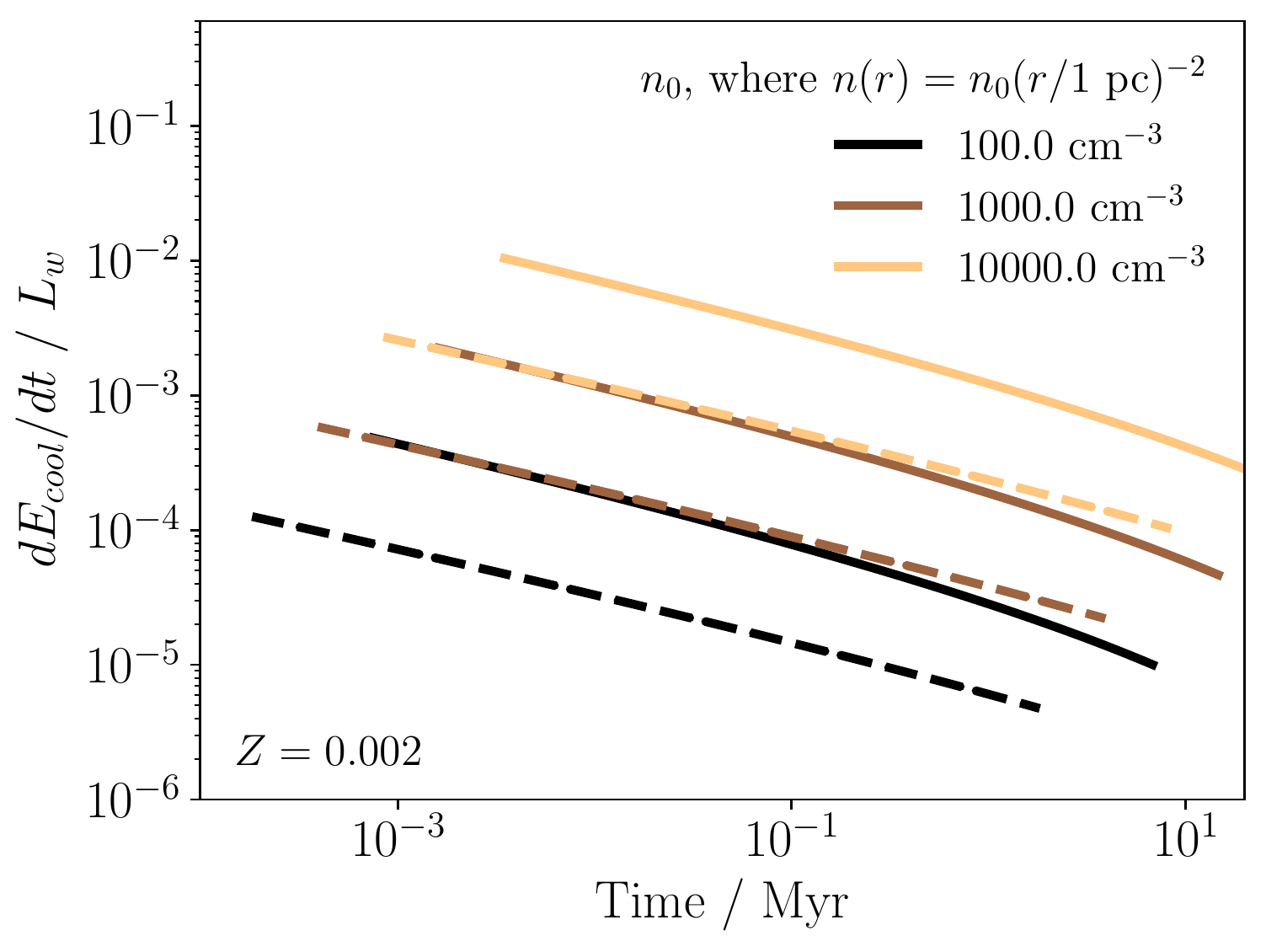}
    \caption{Ratio between cooling rate of the wind bubble $\mathrm{d}E_{cool}/\mathrm{d}t$ and wind luminosity $L_w$ as a function of time, stellar mass and external gas density profile, as discussed in Section \ref{discussion:cooling}. The cooling rate predicted here is far lower than the wind luminosity, particularly at low metallicity, suggesting that the wind bubble should be largely adiabatic under such conditions, provided increased mixing with the background does not occur. \rev{The effect of increasing the stellar mass from 30 to 120 \Msolar is roughly the same as decreasing $n_0$ by a factor of 10.}}
    \label{fig:cooling}
\end{figure*}

In this section we discuss the role of radiative cooling on the wind bubble. To do this, we adopt a similar approach to \cite{MacLow1988}. The total energy loss rate from cooling inside the wind bubble can be written as
\begin{equation}
    \frac{\mathrm{dE_{cool}} }{\mathrm{d} t} = \int_{0}^{r_{lim}}  n_b^2(r) \Lambda(T_b(r),Z) . \Omega r^2 dr
    \label{eqn:coolingrate}
\end{equation}
where the cooling function in the range being considered is given by
\begin{equation}
    \Lambda(T_b(r,t),Z) = 10^{-22}~\mathrm{erg~ cm}^3~\mathrm{s}^{-1}~
    \left(\frac{T_b(r,t)}{10^6~ \mathrm{K}}\right)^{-7/10}
    \left(\frac{Z}{Z_{solar}}\right),
\end{equation}
and $T_b(r)$ and $n_b(r)$ are the temperature and hydrogen number density inside the wind bubble as a function of $r$ (distinct from the external density profile $n(r)$), and $r_{lim} \rightarrow r_w$ is the radius inside which the cooling is considered. \cite{Weaver1977} gives the temperature inside the bubble as
\begin{equation}
    T_b(r,t) = \left(  \frac{P_w r^2_w}{t C} \right)^{2/7} (1-r/r_w)^{2/5} = T_c(t) (1-r/r_w)^{2/5},
\end{equation}
where $T_c$ is the temperature at the centre of the bubble and $C = 6 \times 10^{-7}$ erg s$^{-1}$ cm$^{-1}$ K$^{-7/2}$.
Using the ideal gas equation $P_w = n_b(r) k_B T_b(r)$ together with Equations \ref{eqn:windenergy} and \ref{eqn:windbubbleenergy}, we can find a similar expression for $n_b(r)$,
\begin{equation}
    n_b(r,t) = \frac{P_w^{5/7}}{k_B} \left(  \frac{t C}{ r^2_w} \right)^{2/7} (1-r/r_w)^{-2/5} = n_c(t) (1-r/r_w)^{-2/5},
\end{equation}
where $n_c$ is the particle density at the centre of the bubble.

Since these equations create a singularity in Equation \ref{eqn:coolingrate} at $r_w$, \cite{MacLow1988} pick $r_{lim} = r_w (1 - (T_{cut}/T_c)^{5/2})$ where $T_{cut} = 10^5~$K, arguing that below this temperature the cooling function $\Lambda$ changes shape. In addition, if the wind bubble cooled down to $\sim~10^4~$K, it would reach temperature equilibrium with the photoionised shell outside.

Since $r_{lim}$ is a function of $T_c$, we do not give a fully algebraic solution to Equation \ref{eqn:coolingrate}, but instead plot solutions for it as a function of time in the case where $\omega=2$ in Figure \ref{fig:cooling}. 

We find that for all cases studied, the cooling rate from the wind bubble is 1\% of the wind luminosity or lower, dropping over time as the bubble expands into a lower-density environment, and as such this cooling channel should not affect these calculations. 

There are a few other cooling channels that can become significant. \cite{Fierlinger2016} invoke thermal conduction across the wind bubble interface, as do \cite{Gentry2016} and \cite{ElBadry2019} for supernova remnants. Correct thermal transfer across the contact discontinuity between the wind bubble and the (photoionised) shell is crucial in determining the energy budget of the wind bubble. Further, \cite{Fierlinger2016} argue that cooling in the outer part of the shell as it shock heats against the dense gas should be correctly treated to account for artificial numerical mixing. \rev{Turbulent mixing can also be important under the conditions described in \cite{Tan2021}. Under these conditions, fractal effects on the structure of the wind shell can lead to enhanced cooling (\citealp{Lancaster2021a}, see also \citealp{Fielding2020}), causing a momentum-driven wind bubble with negligible thermal pressure \citep{Silich2013}.} \cite{Rosen2014} argue that cooling is effective in analyses of observed \HII regions, as do \cite{Lancaster2021b} in simulations of wind bubbles in a turbulent density field.

One other cooling mechanism is described in \cite{Arthur2012}, who performed 1D spherically-symmetric hydrodynamic simulations using a uniform external density distribution. They find hotter gas in their simulations than is observed in regions such as the Orion Nebula. They argue that the evaporation of dense embedded \textit{proplyds} (protoplanetary disks affected by feedback) causes mixing with the hot wind bubble, cooling it and slowing its expansion.

We caution that the cooling rate of the wind bubble is very sensitive to the ability for the hot bubble to mix with the cooler, denser gas outside the bubble. More careful hydrodynamic simulations of this phenomenon are needed to constrain the role of cooling further.



\subsection{Gravitational Shell Instability}

Ionising photons can escape if instabilities in the shell cause it to break up sufficiently for low-density channels to form. Gravitational instabilities provide one such channel. For a wind shell moving through a uniform medium ($w=0$), \cite{Elmegreen1994} give the following condition for gravitational instabilities:
\begin{equation}
    \frac{\pi G \rho_0}{3 c_s} > \frac{8^{1/2} v_0}{r_w^2}
\end{equation}
where $c_s$ is the sound speed in the neutral shell and $v_0$ is the speed of the shell in the uniform medium. In this case, the column density of the shell $\sigma_0 = \rho_0 r_w / 3$. For a $\omega=2$ profile, we can use Equation \ref{eqn:massinsider} to give a column density of the shell $\sigma_2 = \rho_0 r_0^2 / r_w$. Modifying this equation to use the new column density, we arrive at a limiting velocity for instabilities of
\begin{equation}
    v_{inst} = \frac{\pi G \rho_0 r_0^2}{8^{1/2} c_s}
\end{equation}
where $v_{w,2} < v_{inst}$ for instability to occur. This gives $v_{inst} \simeq (n_0 / 1000~ \mathrm{cm}^{-3})~$km/s for $c_s = 0.2~$km/s. This is much lower than typical values for $v_{w,2}$ except for the lower mass stars studied here, as in Figure \ref{fig:drdtforw2Metal}, where it is still at least an order of magnitude lower.

Thus we do not expect the wind bubble to experience gravitational instabilities before other forms of photon breakout occur. We further discount Rayleigh-Taylor instabilities since the shell is not accelerating, although even if it were, the growth rate would typically be small compared to the timescales studied in Section \ref{ionised}.

However, this does not discount other forms of instability or interactions with clumps and other aspherical structures in the surrounding medium triggering some form of breakup. There remains a crucial role for detailed 3D hydrodynamic simulations to explore such behaviour.




\subsection{Application to Orion}
\label{discussion:observations}

Recent observations of \rev{nebula M42 in Orion, henceforth referred to as} the Orion Nebula, suggest that it is a wind-driven bubble with a neutral shell travelling at 13 km/s with a derived age of 0.2 Myr \citep{Pabst2019,Pabst2020}. The fact that the shell contains neutral gas implies that the shell traps ionising radiation from the source star, since otherwise the shell would be photoionised. In this section we apply our model to the region to determine whether we predict this behaviour. 

The nebula is powered by the O7V star $\theta^1$ Ori C. This star has an evolutionary mass of $33\pm5$ \Msolar \citep{Balega2014,SimonDiaz2006}. The star is situated close to a dense filament and the nebula is expanding away from the star out to a distance of 4 pc \citep{Guedel2008} in a roughly circular shape on the sky, with the source star near the edge of the nebula. We assume for the purposes of this work that the wind bubble is expanding rapidly into a $\omega=2$ density field, since this is similar to what we find in simulations of young massive stars in clouds similar to this region \citep{Geen2020}.

\begin{figure*}
	\includegraphics[width=2\columnwidth]{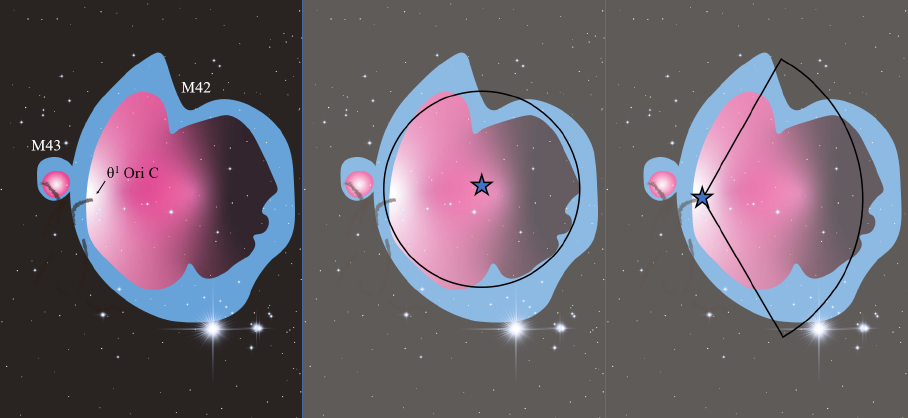}
    \caption{Schematic showing the model used to describe the Orion Nebula (M42). The image on the left shows the location of the principal components of the nebula, including the source of the feedback, star $\theta^1$ Ori C in the Trapezium cluster, at the left edge of the nebula. The nearby nebula M43 is not treated in this work but is included for visual reference. The first model, a sphere centred on the nebula but not the star, is shown in the centre panel. The second model, a cone with its point located on the star with flows moving in a radial direction rightwards in a solid angle of $\pi$, is shown in the right panel. We discuss both models in Section \ref{discussion:observations}.}
    \label{fig:orionschematic}
\end{figure*}

We adopt two models for the initial geometry of the region for the purposes of calculating solutions to the equations in Sections \ref{windmodel} and \ref{ionised}, illustrated with a schematic in Figure \ref{fig:orionschematic}. In the first model, the initial gas mass distribution is centred in the middle of the nebula, with a radius of 2 pc and solid angle subtending the whole sphere around the centre, $\Omega = 4 \pi$. This model is used by authors such as \cite{Arthur2012}, who invoke an \HII region expanding into a uniform medium. 

In addition to this model, we adopt a second model. This is because the position of the star is not at the centre of the nebula, but at one edge bounded by denser gas that constrains the nebula in that direction (see Figure \ref{fig:orionschematic}). To better capture this view of the region, the second model places the star at the corner of a conic section of opening angle $\pi$, reaching out to 4 pc at the edge of the nebula. 

Both models adopt a $\omega=2$ density profile. We consider this to be relatively unlikely in the first model since the remaining neutral dense gas and star are to one edge of the nebula, although we include it for the purposes of comparison with uniform background models in the literature that invoke a spherically symmetric solution with radius 2 pc.

Since the star is rotating only at a few percent of critical velocity \citep{SimonDiaz2006}, we use the non-rotating Geneva tracks to model the region \citep{Ekstrom2012}, with stellar spectra obtained from Starburst99 \citep{Leitherer2014} \rev{(see Table \ref{table:app_starprops_nonrot})}. We give solutions from the 30 and 35 \Msolar tracks to provide estimates of the uncertainty induced by variations in the stellar evolution model for the star.

\cite{Pabst2020} derive from observations masses for the shell around the Orion Nebula between 600 and 2600 \Msolar \citep[or up to 3400 \Msolar with the upper bound on errors given in ][]{Pabst2019}, with an expected value of 1500 \Msolar. The uncertainty depends on the method used to calculate the shell mass and which parts of the shell are considered. In this mass range, $n_0 = 1921^{+1200}_{-1200}~$cm$^{-3}$ if $\Omega = 4 \pi$ and $n_0 = 3842^{+2400}_{-2400}~$cm$^{-3}$ if $\Omega = \pi$.

From this mass range, we can use Equation \ref{eqn:massinsider} to find $\Omega n_0 r_0^2$. \rev{We use these values as inputs to our semi-analytic model in Section$~$\ref{ionised} for the shell expansion. Since the C$^{+}$ emission measured by \cite{Pabst2020} comes from the part of the shell after $r_i$, we use $r_i$ and $\mathrm{d}r_i / \mathrm{d}t$ as the shell radius and expansion velocity, and iterate solutions to the model to find $r_{w,2}$ and $v_{w,2}$. However, we find that $r_i$ is roughly equal to $r_{w,2}$ to within a few percent in all cases listed, so the results should be similar to if we used $r_{w,2}$ and $v_{w,2}$.} 

\begingroup
\renewcommand{\arraystretch}{1.2} 
\begin{table}
    \centering
    \begin{tabular}{cc|ccc}
    \multicolumn{2}{c}{\textbf{Geometry}} & \textbf{Shell velocity} & \textbf{Age} & \textbf{Overflow radius}\\
    $r_i$ / pc & $\Omega$ &  $\mathrm{d}r_i / \mathrm{d}t$ / km/s & t / Myr & $r_{o,2}$ / pc \\
    \hline
    \multicolumn{5}{c}{Stellar mass = 30 \Msolar} \\
    \hline
    2.0 & $4 \pi$ & $9.6^{+4.8}_{-1.7}$ & $0.215^{+0.042}_{-0.067}$ & $44.4^{+40.5}_{-32.4}$
    \\
    \vspace{0.2cm}
    4.0 & $1 \pi$ & $11.1^{+4.7}_{-1.7}$ & $0.359^{+0.065}_{-0.103}$ & $282.2^{+257.0}_{-205.9}$
    \\
    \hline
    \multicolumn{5}{c}{Stellar mass = 35 \Msolar} \\ 
    \hline
    2.0 & $4 \pi$ & $11.7^{+5.6}_{-2.0}$ & $0.176^{+0.033}_{-0.053}$ & $37.7^{+34.3}_{-27.5}$
    \\
    \vspace{0.2cm}        
    4.0 & $1 \pi$ & $13.7^{+5.7}_{-2.1}$ & $0.290^{+0.052}_{-0.083}$ & $239.2^{+217.8}_{-174.5}$
    \\
    \end{tabular}
    \caption{Table showing the results of our model for the conditions in the Orion Nebula. We follow two stellar tracks, one at 30 \Msolar and one at 35 \Msolar in order to capture the uncertainty in the stellar mass measurement for $\theta^1$ Ori C. We further adopt two models for the nebula structure, visualised in Figure \ref{fig:orionschematic}. In the first model, the initial gas mass distribution and the stellar wind and radiation source are centred in the middle of the nebula, with a radius of 2 pc and solid angle subtending the whole sphere around the centre, $\Omega = 4 \pi$. In the second model, the nebula expands in only $\Omega = \pi$ around the star, the rest of the nebula being constrained by denser gas, with the star at the point of the conical volume and the nebula reaching out to 4 pc. We consider the second model to be more representative of the real Orion nebula, where the source of the winds and radiation is to one side of the region. For each model setup, the shell velocity $v_{w,2}$, age of the nebula $t$ and ionising radiation overflow radius $r_{o,2}$ are calculated. The error in the last three quantities comes from the uncertainty in shell mass used - see Section \ref{discussion:observations} for more information. The models predict that the overflow radius of the Orion Nebula is much larger than its current radius, reproducing the observed result that the ionising radiation should be trapped by the shell around the wind bubble.}
    \label{tab:orionmodel}
\end{table}
\endgroup

Table \ref{tab:orionmodel} lists the model setups used and their outcomes. All of the models give velocities and ages that agree with the \cite{Pabst2020} results to within the error given by uncertainties in the shell mass used. \rev{Our preferred model, with the star situated at one end of a \HII region with solid angle $\Omega = \pi$, predicts a slightly older region of around 0.3 Myr, expanding around 1-2 km/s faster, than the model where $\Omega = 4 \pi$ due to the extra time taken to travel 4$~$pc instead of 2$~$pc.}

A few additional processes may affect our result. Internal cooling by proplyds as described in \cite{Arthur2012} may reduce the overall expansion rate of the \HII region. We also note that there are uncertainties in stellar atmosphere and evolution models that may affect our result, as well as uncertainties in the shell masses used that limit a more accurate velocity estimate. Nonetheless, to within the given observational errors, our model is consistent with the observed shell velocity.

The other main result of this comparison is to note that our model predicts that in all cases, the Orion Nebula has an overflow radius well above its current radius, suggesting that the ionising radiation should be easily trapped by the shell and cause the Orion Nebula to be a wind-driven structure. \rev{The overflow radius is significantly larger in the case where $\Omega = \pi$ since the wind is compressed into a smaller volume, increasing its relative pressure and hence reducing the likelihood of overflow.}

We note that the lack of 3D information in the model limits our ability to predict the exact behaviour of the shell when turbulence and instabilities occur. A better constrained model would require closer comparison to the observations, perhaps using 3D hydrodynamic simulations to account for the role of instabilities, turbulence, and other 3D effects in the system. Rather, the purpose of this work is to establish the feasibility for such behaviour of wind bubbles in trapping ionising radiation and preventing photoionised champagne flows as observed in systems like the Orion Nebula.

\rev{There are a number of possible reasons why simulations to date do not reproduce systems behaving in this way, such as the Orion Nebula, in addition to the cooling arguments made in Section \ref{discussion:cooling}. One is that they lack the spatial resolution to properly capture the dense shell, and hence the ionising photon trapping. Another is the role of magnetic fields, which act to prevent the breakup of extended structures in molecular clouds \citep{Hennebelle2013}, and hence the fragmentation of feedback-driven shells \citep{Geen2015b}. Finally, the precise conditions of feedback in the cloud may simply determine whether the wind bubble is surrounded by sufficient dense cloud material to prevent a breakout into the diffuse external medium. Further detailed comparison is needed to determine whether the discrepancies between conclusions from observations and numerical theory can be resolved.}

\section{Conclusions}

In this paper we study the expansion of wind bubbles in power law density fields around young massive stars, with a particular focus on ``singular isothermal sphere'' density fields $(\rho \propto r^{-2})$ as predicted to exist around young stars. 

The goal of this paper is to a) derive analytic solutions for wind bubble expansion in non-uniform density fields, b) calculate at what point ionising radiation can escape such a system and ionise the surrounding medium and c) determine at what point effects not included in this model grow to require more detailed numerical models. The reason for undertaking this study is to reconcile certain observations, which suggest feedback around young massive stars is driven principally by winds and not ionising radiation, and theory, which has often argued that winds are not dynamically important.

We find that adiabatic wind bubbles expanding into density fields where $\rho \propto r^{-2}$ do so at a constant rate $v_{w,2}$, which depends on the wind luminosity $L_w$ and characteristic cloud density $\rho_0$. Radiation pressure contributes only a small amount to the expansion of the \HII region. In denser cloud environments, ionising radiation cannot overflow the shell around the wind bubble and escape into the surrounding medium is prevented until the shell reaches a size of several parsecs and an age of $\gtrsim1~$Myr. A more likely scenario is that the conditions in the external medium change to allow shell fragmentation and ionising photon escape. 

Radiative cooling, magnetic fields, gravity and instabilities do not appear to have a strong influence on the solution, although gravity can become a larger perturbation around less massive O stars in dense clouds.

Our model provides a framework for understanding the behaviour of the Orion Nebula. Our preferred model is one in which the 30-35$~$\Msolar star $\theta^1$ Ori C sits at one edge of the nebula, as is observed, and the nebula expands into a singular isothermal sphere over a solid angle of $\sim \pi$, constrained by denser gas in all other directions. Using this model, we recover an expansion velocity consistent with the observed shell velocity of 13 km/s to within the errors given by the observational studies. We also explain that the wind-driven shell should easily trap ionising radiation under the conditions found in Orion, as observed.

The solution found in this work is likely to break down as the wind bubble enters the wider structured cloud environment. This highlights the need for detailed 3D radiative magnetohydrodynamic simulations. At the same time, sufficient resolution and care with sub-grid feedback injection recipes is crucial in correctly capturing the interaction between stellar winds, radiation and the star-forming environment. To reproduce this \rev{phenomenon}, simulations should correctly capture the initial wind shell formation before a champagne flow occurs.

In summary, stellar winds around young massive stars should not be discounted from a theoretical perspective. Careful interaction between analytic theory, numerical simulations and observations are needed to determine precisely how massive stars shape their environment.

\section*{Acknowledgements}

\rev{The authors would like to thank the anonymous referee for comments that helped greatly improve the manuscript.} The authors would like to thank Xander Tielens, Cornelia Pabst, and Eric Pellegrini for useful discussions during the writing of this paper. SG acknowledges support from a NOVA grant for the theory of massive star formation.

\section*{Data Availability}

The code and data products used in this work are stored according to the Anton Pannekoek Institute for Astronomy Research Data Management plan, and can be found via DOI 10.5281/zenodo.5646533.



\bibliographystyle{mnras}
\bibliography{samgeen} 

\begin{thebibliography}{}
\makeatletter
\relax
\def\mn@urlcharsother{\let\do\@makeother \do\$\do\&\do\#\do\^\do\_\do\%\do\~}
\def\mn@doi{\begingroup\mn@urlcharsother \@ifnextchar [ {\mn@doi@}
  {\mn@doi@[]}}
\def\mn@doi@[#1]#2{\def\@tempa{#1}\ifx\@tempa\@empty \href
  {http://dx.doi.org/#2} {doi:#2}\else \href {http://dx.doi.org/#2} {#1}\fi
  \endgroup}
\def\mn@eprint#1#2{\mn@eprint@#1:#2::\@nil}
\def\mn@eprint@arXiv#1{\href {http://arxiv.org/abs/#1} {{\tt arXiv:#1}}}
\def\mn@eprint@dblp#1{\href {http://dblp.uni-trier.de/rec/bibtex/#1.xml}
  {dblp:#1}}
\def\mn@eprint@#1:#2:#3:#4\@nil{\def\@tempa {#1}\def\@tempb {#2}\def\@tempc
  {#3}\ifx \@tempc \@empty \let \@tempc \@tempb \let \@tempb \@tempa \fi \ifx
  \@tempb \@empty \def\@tempb {arXiv}\fi \@ifundefined
  {mn@eprint@\@tempb}{\@tempb:\@tempc}{\expandafter \expandafter \csname
  mn@eprint@\@tempb\endcsname \expandafter{\@tempc}}}

\bibitem[\protect\citeauthoryear{Agertz, Kravtsov, Leitner  \& Gnedin}{Agertz
  et~al.}{2013}]{Agertz2013}
Agertz O.,  Kravtsov A.~V.,  Leitner S.~N.,   Gnedin N.~Y.,  2013, \mn@doi [The
  Astrophysical Journal] {10.1088/0004-637X/770/1/25}, 770, 25

\bibitem[\protect\citeauthoryear{Ali, Harries  \& Douglas}{Ali
  et~al.}{2018}]{Ali2018}
Ali A.,  Harries T.~J.,   Douglas T.~A.,  2018, \mn@doi [Monthly Notices of the
  Royal Astronomical Society] {10.1093/mnras/sty1001}, 477, 5422

\bibitem[\protect\citeauthoryear{Arthur}{Arthur}{2012}]{Arthur2012}
Arthur S.~J.,  2012, \mn@doi [Monthly Notices of the Royal Astronomical
  Society] {10.1111/j.1365-2966.2011.20388.x}, 421, 1283

\bibitem[\protect\citeauthoryear{Avedisova}{Avedisova}{1972}]{Avedisova1972}
Avedisova V.~S.,  1972, Soviet Astronomy, 15

\bibitem[\protect\citeauthoryear{Baker \& Menzel}{Baker \&
  Menzel}{1938}]{Baker1938}
Baker J.~G.,  Menzel D.~H.,  1938, \mn@doi [The Astrophysical Journal]
  {10.1086/143959}, 88, 52

\bibitem[\protect\citeauthoryear{Balega, Chentsov, Leushin, Rzaev  \&
  Weigelt}{Balega et~al.}{2014}]{Balega2014}
Balega Y.~Y.,  Chentsov E.,  Leushin V.,  Rzaev A.~K.,   Weigelt G.,  2014,
  \mn@doi [Astrophysical Bulletin] {10.1134/S1990341314010052}, 69, 46

\bibitem[\protect\citeauthoryear{Bate}{Bate}{2012}]{Bate2012}
Bate M.~R.,  2012, \mn@doi [Monthly Notices of the Royal Astronomical Society]
  {10.1111/j.1365-2966.2011.19955.x}, 419, 3115

\bibitem[\protect\citeauthoryear{Calder{\'{o}}n, Cuadra, Schartmann, Burkert,
  Prieto  \& Russell}{Calder{\'{o}}n et~al.}{2020}]{Calderon2020}
Calder{\'{o}}n D.,  Cuadra J.,  Schartmann M.,  Burkert A.,  Prieto J.,
  Russell C. M.~P.,  2020, Monthly Notices of the Royal Astronomical Society,
  Advance Access, 493, 447

\bibitem[\protect\citeauthoryear{Capriotti \& Kozminski}{Capriotti \&
  Kozminski}{2001}]{Capriotti2001}
Capriotti E.,  Kozminski J.,  2001, \mn@doi [Publications of the Astronomical
  Society of the Pacific] {10.1086/320809}, 113, 677

\bibitem[\protect\citeauthoryear{Castor, Weaver  \& McCray}{Castor
  et~al.}{1975}]{Castor1975}
Castor J.,  Weaver R.,   McCray R.,  1975, \mn@doi [The Astrophysical Journal]
  {10.1086/181908}, 200, L107

\bibitem[\protect\citeauthoryear{Comeron}{Comeron}{1997}]{Comeron1997}
Comeron F.,  1997, Astronomy \& Astrophysics, 326, 1195

\bibitem[\protect\citeauthoryear{Crutcher}{Crutcher}{2012}]{Crutcher2012}
Crutcher R.~M.,  2012, \mn@doi [Annual Review of Astronomy and Astrophysics]
  {10.1146/annurev-astro-081811-125514}, 50, 29

\bibitem[\protect\citeauthoryear{Dale, Ercolano  \& Bonnell}{Dale
  et~al.}{2012}]{Dale2012}
Dale J.~E.,  Ercolano B.,   Bonnell I.~A.,  2012, \mn@doi [Monthly Notices of
  the Royal Astronomical Society] {10.1111/j.1365-2966.2012.21205.x}, 424, 377

\bibitem[\protect\citeauthoryear{Dale, Ngoumou, Ercolano  \& Bonnell}{Dale
  et~al.}{2014}]{Dale2014}
Dale J.~E.,  Ngoumou J.,  Ercolano B.,   Bonnell I.~A.,  2014, \mn@doi [Monthly
  Notices of the Royal Astronomical Society] {10.1093/mnras/stu816}, 442, 694

\bibitem[\protect\citeauthoryear{Decataldo, Lupi, Ferrara, Pallottini  \&
  Fumagalli}{Decataldo et~al.}{2020}]{Decataldo2020}
Decataldo D.,  Lupi A.,  Ferrara A.,  Pallottini A.,   Fumagalli M.,  2020,
  \mn@doi [Monthly Notices of the Royal Astronomical Society]
  {10.1093/mnras/staa2326}, 497, 4718

\bibitem[\protect\citeauthoryear{Draine}{Draine}{2011}]{Draine2011}
Draine B.~T.,  2011, \mn@doi [The Astrophysical Journal]
  {10.1088/0004-637X/732/2/100}, 732, 100

\bibitem[\protect\citeauthoryear{Dunne, Chu, Chen, Lowry, Townsley, Gruendl,
  Guerrero  \& Rosado}{Dunne et~al.}{2003}]{Dunne2003}
Dunne B.~C.,  Chu Y.~H.,  Chen C. H.~R.,  Lowry J.~D.,  Townsley L.,  Gruendl
  R.~A.,  Guerrero M.~A.,   Rosado M.,  2003, \mn@doi [The Astrophysical
  Journal] {10.1086/375010}, 590, 306

\bibitem[\protect\citeauthoryear{Dyson \& Williams}{Dyson \&
  Williams}{1980}]{Dyson1980}
Dyson J.~E.,  Williams D.~A.,  1980, {Physics of the interstellar medium}.
New York, Halsted Press

\bibitem[\protect\citeauthoryear{Ekstr{\"{o}}m et~al.,}{Ekstr{\"{o}}m
  et~al.}{2012}]{Ekstrom2012}
Ekstr{\"{o}}m S.,  et~al., 2012, \mn@doi [Astronomy \& Astrophysics]
  {10.1051/0004-6361/201117751}, 537, A146

\bibitem[\protect\citeauthoryear{El-Badry, Ostriker, Kim, Quataert  \&
  Weisz}{El-Badry et~al.}{2019}]{ElBadry2019}
El-Badry K.,  Ostriker E.~C.,  Kim C.-G.,  Quataert E.,   Weisz D.~R.,  2019,
  \mn@doi [Monthly Notices of the Royal Astronomical Society]
  {10.1093/mnras/stz2773}, 490, 1961

\bibitem[\protect\citeauthoryear{Elmegreen}{Elmegreen}{1994}]{Elmegreen1994}
Elmegreen B.~G.,  1994, \mn@doi [The Astrophysical Journal] {10.1086/174147},
  427, 384

\bibitem[\protect\citeauthoryear{Fielding, Ostriker, Bryan  \& Jermyn}{Fielding
  et~al.}{2020}]{Fielding2020}
Fielding D.~B.,  Ostriker E.~C.,  Bryan G.~L.,   Jermyn A.~S.,  2020, \mn@doi
  [Astrophysical Journal Letters] {10.3847/2041-8213/ab8d2c}, 894, L24

\bibitem[\protect\citeauthoryear{Fierlinger, Burkert, Ntormousi, Fierlinger,
  Schartmann, Ballone, Krause  \& Diehl}{Fierlinger
  et~al.}{2016}]{Fierlinger2016}
Fierlinger K.~M.,  Burkert A.,  Ntormousi E.,  Fierlinger P.,  Schartmann M.,
  Ballone A.,  Krause M. G.~H.,   Diehl R.,  2016, \mn@doi [Monthly Notices of
  the Royal Astronomical Society] {10.1093/mnras/stv2699}, 456, 710

\bibitem[\protect\citeauthoryear{Franco, Tenorio-Tagle  \& Bodenheimer}{Franco
  et~al.}{1990}]{Franco1990}
Franco J.,  Tenorio-Tagle G.,   Bodenheimer P.,  1990, \mn@doi [The
  Astrophysical Journal] {10.1086/168300}, 349, 126

\bibitem[\protect\citeauthoryear{Gallegos-Garcia, Burkhart, Rosen, Naiman  \&
  Ramirez-Ruiz}{Gallegos-Garcia et~al.}{2020}]{GallegosGarcia2020}
Gallegos-Garcia M.,  Burkhart B.,  Rosen A.~L.,  Naiman J.~P.,   Ramirez-Ruiz
  E.,  2020, \mn@doi [Astrophysical Journal Letters]
  {10.3847/2041-8213/ababae}, 899, L30

\bibitem[\protect\citeauthoryear{Gatto et~al.,}{Gatto et~al.}{2017}]{Gatto2017}
Gatto A.,  et~al., 2017, \mn@doi [Monthly Notices of the Royal Astronomical
  Society] {10.1093/mnras/stw3209}, 466, 1903

\bibitem[\protect\citeauthoryear{Geen, Hennebelle, Tremblin  \& Rosdahl}{Geen
  et~al.}{2015}]{Geen2015b}
Geen S.,  Hennebelle P.,  Tremblin P.,   Rosdahl J.,  2015, \mn@doi [Monthly
  Notices of the Royal Astronomical Society] {10.1093/mnras/stv2272}, 454, 4484

\bibitem[\protect\citeauthoryear{Geen, Pellegrini, Bieri  \& Klessen}{Geen
  et~al.}{2020}]{Geen2019}
Geen S.,  Pellegrini E.,  Bieri R.,   Klessen R.,  2020, \mn@doi [Monthly
  Notices of the Royal Astronomical Society] {10.1093/mnras/stz3491}, 492, 915

\bibitem[\protect\citeauthoryear{Geen, Bieri, Rosdahl  \& de Koter}{Geen
  et~al.}{2021}]{Geen2020}
Geen S.,  Bieri R.,  Rosdahl J.,   de Koter A.,  2021, Monthly Notices of the
  Royal Astronomical Society, 501, 1352

\bibitem[\protect\citeauthoryear{Gentry, Krumholz, Dekel  \& Madau}{Gentry
  et~al.}{2017}]{Gentry2016}
Gentry E.~S.,  Krumholz M.~R.,  Dekel A.,   Madau P.,  2017, Monthly Notices of
  the Royal Astronomical Society, 465, 2471

\bibitem[\protect\citeauthoryear{Groenewegen, Lamers  \& Pauldrach}{Groenewegen
  et~al.}{1989}]{Groenewegen1989}
Groenewegen M.,  Lamers H.,   Pauldrach A.,  1989, Astronomy \& Astrophysics,
  221, 78

\bibitem[\protect\citeauthoryear{Grudi{\'{c}}, Kruijssen,
  Faucher-Gigu{\`{e}}re, Hopkins, Ma, Quataert  \& Boylan-Kolchin}{Grudi{\'{c}}
  et~al.}{2020}]{Grudic2020}
Grudi{\'{c}} M.~Y.,  Kruijssen J.~D.,  Faucher-Gigu{\`{e}}re C.-A.,  Hopkins
  P.~F.,  Ma X.,  Quataert E.,   Boylan-Kolchin M.,  2020, arXiv e-prints, p.
  arXiv:2008.04453

\bibitem[\protect\citeauthoryear{Guedel, Briggs, Montmerle, Audard, Rebull  \&
  Skinner}{Guedel et~al.}{2008}]{Guedel2008}
Guedel M.,  Briggs K.~R.,  Montmerle T.,  Audard M.,  Rebull L.,   Skinner
  S.~L.,  2008, \mn@doi [Science] {10.1126/science.1149926}, 319, 309

\bibitem[\protect\citeauthoryear{Haid, Walch, Seifried, W{\"{u}}nsch, Dinnbier
  \& Naab}{Haid et~al.}{2018}]{Haid2018}
Haid S.,  Walch S.,  Seifried D.,  W{\"{u}}nsch R.,  Dinnbier F.,   Naab T.,
  2018, \mn@doi [Monthly Notices of the Royal Astronomical Society]
  {10.1093/mnras/sty1315}, 478, 4799

\bibitem[\protect\citeauthoryear{Hennebelle}{Hennebelle}{2013}]{Hennebelle2013}
Hennebelle P.,  2013, \mn@doi [Astronomy \& Astrophysics]
  {10.1051/0004-6361/201321292}, 556, A153

\bibitem[\protect\citeauthoryear{Howarth, Siebert, Hussain  \& Prinja}{Howarth
  et~al.}{1997}]{Howarth1997}
Howarth I.~D.,  Siebert K.~W.,  Hussain G.~A.,   Prinja R.~K.,  1997, \mn@doi
  [Monthly Notices of the Royal Astronomical Society]
  {10.1093/mnras/284.2.265}, 284, 265

\bibitem[\protect\citeauthoryear{Kahn}{Kahn}{1954}]{KahnF.D.1954}
Kahn F.~D.,  1954, Bulletin of the Astronomical Institutes of the Netherlands,
  12, 187

\bibitem[\protect\citeauthoryear{Kim, Kim  \& Ostriker}{Kim
  et~al.}{2016}]{Kim2016}
Kim J.-G.,  Kim W.-T.,   Ostriker E.~C.,  2016, \mn@doi [The Astrophysical
  Journal] {10.3847/0004-637X/819/2/137}, 819, 23

\bibitem[\protect\citeauthoryear{Krumholz \& Matzner}{Krumholz \&
  Matzner}{2009}]{Krumholz2009}
Krumholz M.~R.,  Matzner C.~D.,  2009, \mn@doi [The Astrophysical Journal]
  {10.1088/0004-637X/703/2/1352}, 703, 1352

\bibitem[\protect\citeauthoryear{Kuiper \& Hosokawa}{Kuiper \&
  Hosokawa}{2018}]{Kuiper2018}
Kuiper R.,  Hosokawa T.,  2018, Astronomy \& Astrophysics, 616, 22

\bibitem[\protect\citeauthoryear{Lamers, Leitherer, Lamers  \&
  Leitherer}{Lamers et~al.}{1993}]{Lamers1993}
Lamers H. J. G. L.~M.,  Leitherer C.,  Lamers H. J. G. L.~M.,   Leitherer C.,
  1993, \mn@doi [The Astrophysical Journal] {10.1086/172960}, 412, 771

\bibitem[\protect\citeauthoryear{Lancaster, Ostriker, Kim  \& Kim}{Lancaster
  et~al.}{2021a}]{Lancaster2021a}
Lancaster L.,  Ostriker E.~C.,  Kim J.-G.,   Kim C.-G.,  2021a, \mn@doi [The
  Astrophysical Journal] {10.3847/1538-4357/abf8ab}, 914, 89

\bibitem[\protect\citeauthoryear{Lancaster, Ostriker, Kim  \& Kim}{Lancaster
  et~al.}{2021b}]{Lancaster2021b}
Lancaster L.,  Ostriker E.~C.,  Kim J.-G.,   Kim C.-G.,  2021b, \mn@doi [The
  Astrophysical Journal] {10.3847/1538-4357/abf8ac}, 914, 90

\bibitem[\protect\citeauthoryear{Lee \& Hennebelle}{Lee \&
  Hennebelle}{2018}]{Lee2018}
Lee Y.-N.,  Hennebelle P.,  2018, \mn@doi [Astronomy \& Astrophysics]
  {10.1051/0004-6361/201731522}, 611, A88

\bibitem[\protect\citeauthoryear{Leitherer, Ekstr{\"{o}}m, Meynet, Schaerer,
  Agienko  \& Levesque}{Leitherer et~al.}{2014}]{Leitherer2014}
Leitherer C.,  Ekstr{\"{o}}m S.,  Meynet G.,  Schaerer D.,  Agienko K.~B.,
  Levesque E.~M.,  2014, \mn@doi [The Astrophysical Journal Supplement Series]
  {10.1088/0067-0049/212/1/14}, 212, 14

\bibitem[\protect\citeauthoryear{{Mac Low} \& McCray}{{Mac Low} \&
  McCray}{1988}]{MacLow1988}
{Mac Low} M.-M.,  McCray R.,  1988, \mn@doi [The Astrophysical Journal]
  {10.1086/165936}, 324, 776

\bibitem[\protect\citeauthoryear{Mart{\'{i}}nez-Gonz{\'{a}}lez, Silich  \&
  Tenorio-Tagle}{Mart{\'{i}}nez-Gonz{\'{a}}lez
  et~al.}{2014}]{Martinez-Gonzalez2014}
Mart{\'{i}}nez-Gonz{\'{a}}lez S.,  Silich S.,   Tenorio-Tagle G.,  2014,
  \mn@doi [Astrophysical Journal] {10.1088/0004-637X/785/2/164}, 785, 164

\bibitem[\protect\citeauthoryear{Massey, Puls, Pauldrach, Bresolin, Kudritzki
  \& Simon}{Massey et~al.}{2005}]{Massey2005}
Massey P.,  Puls J.,  Pauldrach A.,  Bresolin F.,  Kudritzki R.~P.,   Simon T.,
   2005, \mn@doi [Astrophysical Journal] {10.1086/430417}, 627, 477

\bibitem[\protect\citeauthoryear{Mathews}{Mathews}{1967}]{Mathews1967}
Mathews W.~G.,  1967, \mn@doi [The Astrophysical Journal] {10.1086/149087},
  147, 965

\bibitem[\protect\citeauthoryear{Murray, Quataert  \& Thompson}{Murray
  et~al.}{2010}]{Murray2010}
Murray N.,  Quataert E.,   Thompson T.~A.,  2010, \mn@doi [The Astrophysical
  Journal] {10.1088/0004-637X/709/1/191}, 709, 191

\bibitem[\protect\citeauthoryear{Oort \& Spitzer}{Oort \&
  Spitzer}{1955}]{Oort1955}
Oort J.~H.,  Spitzer L.,  1955, \mn@doi [The Astrophysical Journal]
  {10.1086/145958}, 121, 6

\bibitem[\protect\citeauthoryear{Pabst et~al.,}{Pabst et~al.}{2019}]{Pabst2019}
Pabst C.,  et~al., 2019, \mn@doi [Nature] {10.1038/s41586-018-0844-1}, 565, 618

\bibitem[\protect\citeauthoryear{Pabst et~al.,}{Pabst et~al.}{2020}]{Pabst2020}
Pabst C. H.~M.,  et~al., 2020, \mn@doi [Astronomy \& Astrophysics]
  {10.1051/0004-6361/202037560}, 639, A2

\bibitem[\protect\citeauthoryear{Pellegrini et~al.,}{Pellegrini
  et~al.}{2007}]{Pellegrini2007}
Pellegrini E.~W.,  et~al., 2007, \mn@doi [The Astrophysical Journal]
  {10.1086/511258}, 658, 1119

\bibitem[\protect\citeauthoryear{Pellegrini, Baldwin, Ferland, Shaw  \&
  Heathcote}{Pellegrini et~al.}{2009}]{Pellegrini2009}
Pellegrini E.~W.,  Baldwin J.~A.,  Ferland G.~J.,  Shaw G.,   Heathcote S.,
  2009, \mn@doi [The Astrophysical Journal] {10.1088/0004-637X/693/1/285}, 693,
  285

\bibitem[\protect\citeauthoryear{Pellegrini, Rahner, Reissl, Glover, Klessen,
  Rousseau-Nepton  \& Herrera-Camus}{Pellegrini et~al.}{2020}]{Pellegrini2020}
Pellegrini E.~W.,  Rahner D.,  Reissl S.,  Glover S. C.~O.,  Klessen R.~S.,
  Rousseau-Nepton L.,   Herrera-Camus R.,  2020, Monthly Notices of the Royal
  Astronomical Society, 496, pp339

\bibitem[\protect\citeauthoryear{Rahner, Pellegrini, Glover  \& Klessen}{Rahner
  et~al.}{2017}]{Rahner2017}
Rahner D.,  Pellegrini E.~W.,  Glover S. C.~O.,   Klessen R.~S.,  2017, Monthly
  Notices of the Royal Astronomical Society, 470, 4453

\bibitem[\protect\citeauthoryear{Rogers \& Pittard}{Rogers \&
  Pittard}{2013}]{Rogers2013}
Rogers H.,  Pittard J.~M.,  2013, \mn@doi [Monthly Notices of the Royal
  Astronomical Society] {10.1093/mnras/stt255}, 431, 1337

\bibitem[\protect\citeauthoryear{Rosdahl, Blaizot, Aubert, Stranex  \&
  Teyssier}{Rosdahl et~al.}{2013}]{Rosdahl2013}
Rosdahl J.,  Blaizot J.,  Aubert D.,  Stranex T.,   Teyssier R.,  2013, \mn@doi
  [Monthly Notices of the Royal Astronomical Society] {10.1093/mnras/stt1722},
  436, 2188

\bibitem[\protect\citeauthoryear{Rosen, Lopez, Krumholz  \& Ramirez-Ruiz}{Rosen
  et~al.}{2014}]{Rosen2014}
Rosen A.~L.,  Lopez L.~A.,  Krumholz M.~R.,   Ramirez-Ruiz E.,  2014, \mn@doi
  [Monthly Notices of the Royal Astronomical Society] {10.1093/mnras/stu1037},
  442, 2701

\bibitem[\protect\citeauthoryear{Shu, Lizano, Galli, Canto  \& Laughlin}{Shu
  et~al.}{2002}]{Shu2002}
Shu F.,  Lizano S.,  Galli D.,  Canto J.,   Laughlin G.,  2002, \mn@doi [The
  Astrophysical Journal] {10.1086/343859}, 580, 969

\bibitem[\protect\citeauthoryear{Silich \& Tenorio-Tagle}{Silich \&
  Tenorio-Tagle}{2013}]{Silich2013}
Silich S.,  Tenorio-Tagle G.,  2013, \mn@doi [The Astrophysical Journal]
  {10.1088/0004-637X/765/1/43}, 765

\bibitem[\protect\citeauthoryear{Sim{\'{o}}n-D{\'{i}}az, Herrero, Esteban  \&
  Najarro}{Sim{\'{o}}n-D{\'{i}}az et~al.}{2006}]{SimonDiaz2006}
Sim{\'{o}}n-D{\'{i}}az S.,  Herrero A.,  Esteban C.,   Najarro F.,  2006,
  \mn@doi [Astronomy \& Astrophysics] {10.1051/0004-6361:20053066}, 448, 351

\bibitem[\protect\citeauthoryear{Spitzer}{Spitzer}{1978}]{SpitzerLyman1978}
Spitzer L.,  1978, {Physical processes in the interstellar medium}.
New York Wiley-Interscience

\bibitem[\protect\citeauthoryear{Tan, Oh  \& Gronke}{Tan
  et~al.}{2021}]{Tan2021}
Tan B.,  Oh S.~P.,   Gronke M.,  2021, \mn@doi [Monthly Notices of the Royal
  Astronomical Society] {10.1093/mnras/stab053}, 502, 3179

\bibitem[\protect\citeauthoryear{Tenorio-Tagle}{Tenorio-Tagle}{1979}]{TenorioTagle1979}
Tenorio-Tagle G.,  1979, Astronomy and Astrophysics, 71, 59

\bibitem[\protect\citeauthoryear{Verdolini et~al.,}{Verdolini
  et~al.}{2013}]{Verdolini2013}
Verdolini S.,  et~al., 2013, \mn@doi [The Astrophysical Journal]
  {10.1088/0004-637X/769/1/12}, 769, 12

\bibitem[\protect\citeauthoryear{Vink, Muijres, Anthonisse, de Koter, Graefener
   \& Langer}{Vink et~al.}{2011}]{Vink2011}
Vink J.~S.,  Muijres L.~E.,  Anthonisse B.,  de Koter A.,  Graefener G.,
  Langer N.,  2011, \mn@doi [Astronomy \& Astrophysics]
  {10.1051/0004-6361/201116614}, 531, 132

\bibitem[\protect\citeauthoryear{Wall, McMillan, {Mac Low}, Klessen  \&
  Zwart}{Wall et~al.}{2019}]{Wall2019}
Wall J.~E.,  McMillan S. L.~W.,  {Mac Low} M.-M.,  Klessen R.~S.,   Zwart
  S.~P.,  2019, Astrophysical Journal, 887

\bibitem[\protect\citeauthoryear{Wall, {Mac Low}, McMillan, Klessen, {Portegies
  Zwart}  \& Pellegrino}{Wall et~al.}{2020}]{Wall2020}
Wall J.~E.,  {Mac Low} M.-M.,  McMillan S.~L.,  Klessen R.~S.,  {Portegies
  Zwart} S.,   Pellegrino A.,  2020, \mn@doi [Astrophysical Journal]
  {10.3847/1538-4357/abc011}, 904, 192

\bibitem[\protect\citeauthoryear{Weaver, McCray, Castor, Shapiro  \&
  Moore}{Weaver et~al.}{1977}]{Weaver1977}
Weaver R.,  McCray R.,  Castor J.,  Shapiro P.,   Moore R.,  1977, \mn@doi [The
  Astrophysical Journal] {10.1086/155692}, 218, 377

\bibitem[\protect\citeauthoryear{Yeh \& Matzner}{Yeh \&
  Matzner}{2012}]{Yeh2012}
Yeh S. C.~C.,  Matzner C.~D.,  2012, \mn@doi [The Astrophysical Journal]
  {10.1088/0004-637X/757/2/108}, 757, 18

\bibitem[\protect\citeauthoryear{Zamora-Avil{\'{e}}s
  et~al.,}{Zamora-Avil{\'{e}}s et~al.}{2019}]{Zamora-Aviles2019}
Zamora-Avil{\'{e}}s M.,  et~al., 2019, Monthly Notices of the Royal
  Astronomical Society, 487, 2200

\makeatother
\end{thebibliography}




\appendix

\section{Calculation of the Energy Partition in the Wind Bubble}
\label{appendix:analyticenergypartition}

\rev{In this Appendix we derive the fraction of energy from the wind that remains in the wind bubble, as opposed to being used to do work on the dense shell. This is a generalised re-derivation of Section II of \cite{Weaver1977}. }

\rev{We begin with a dimensional analysis which establishes the power law relationships between the radius of the wind bubble $r_w$, the wind luminosity $L_w$, the background density at $r_w$, $\rho(r_w) = \rho_0 (r_w/r_0)^{-\omega}$ and time $t$:
}
\begin{equation}
    r_w^5(t) \rho(r_w) L_w^{-1}  t^{-3} \propto 1 
    \label{eqn:app_dimensionalanalysis}
\end{equation}
\rev{and thus}
\begin{equation}
    r_w(t) \propto \left( L_w  t^{3} r_0^{-\omega} \rho_0^{-1} \right)^{1/(5-\omega)}.
    \label{eqn:app_dimensionalanalysis2}
\end{equation}
\rev{We further assume a constant partition of the energy added by the wind to the bubble }
\begin{equation}
    E_b = f_{b} L_w t
    \label{eqn:app_energyfraction}
\end{equation}
\rev{where $f_b$ is a fraction to be derived.}

\rev{\cite{Weaver1977} divide the wind bubble into three phases. The first, from $0 \leq r \leq R_1$, is the free-streaming stellar wind. At $R_1$, the wind shocks against the surrounding medium, and from $R_1 \leq r \leq R_c$ is the shocked stellar wind. Finally, from $R_c \leq r \leq R_2$ is the shocked interstellar material, which becomes a dense shell once radiative cooling becomes efficient and the internal pressure reduces. Due to the lower density in the region $R_1 \leq r \leq R_c$, this volume remains roughly adiabatic until later times.}

\rev{We adopt the hydrostatic approximation as in \cite{Weaver1977} for the region $R_1 \leq r \leq R_c$ due to the typically high sound speeds and short sound-crossing times. Thus the internal pressure $P(r,t) = P(t) = P(R_c)$. }

\rev{The continuity equation for mass is given by}
\begin{equation}
\frac{D \rho}{D t} + \frac{\rho}{r^2}\frac{\partial (r^2 v)}{\partial r} = 0.
\label{eqn:app_continuity}
\end{equation}
\rev{This is the standard equation for mass conservation in spherical coordinates, assuming all flows occur in radial directions, where the material derivative $D/D t$ is defined as}
\begin{equation}
\frac{D }{D t} \equiv \frac{\partial }{\partial t} + v \frac{\partial }{\partial r}.
\end{equation}

\rev{We further invoke the energy conservation equation for the adiabatic flow}
\begin{equation}
    \frac{D (P \rho^{-\gamma})}{D t} = 0 
    \label{eqn:app_conservationofenergy}
\end{equation}
\rev{where $\gamma = 5/3$, i.e. the gas is monatomic.}

\rev{Expanding Equation} \ref{eqn:app_conservationofenergy} gives us
\begin{equation}
    \frac{D \rho}{D t} =  \frac{\rho}{P \gamma}\frac{D P}{D t},
\end{equation}
\rev{and hence we can use Equation} \ref{eqn:app_continuity} to write
\begin{equation}
    \frac{1}{r^2}\frac{\partial (r^2 v)}{\partial r} = \frac{1}{P \gamma}\frac{DP}{D t} .
    \label{eqn:app_continuity2}
\end{equation}

\rev{To solve the right hand side of this equation, we note that the pressure in the bubble is the energy divided by the volume, and thus $P \propto L_w t / r^3$. Using Equation \ref{eqn:app_dimensionalanalysis2}, we can write}
\begin{equation}
P \propto t^{-(4+\omega)/(5 - \omega)}    
\end{equation}
\rev{and so }
\begin{equation}
\frac{1}{P \gamma}\frac{D P}{D t} = \frac{1}{P \gamma}\frac{\partial P}{\partial t} = -\frac{3}{5}.\frac{4+\omega}{5 - \omega} \frac{1}{t}.
\end{equation}

\rev{Substituting this into Equation \ref{eqn:app_continuity2} and integrating, we can write}
\begin{equation}
    v(r,t) = \frac{f(t)}{r^2} +  \frac{4 + \omega}{25 - 5\omega} \frac{r}{t}.
\end{equation}
\rev{By differentiating Equation \ref{eqn:app_dimensionalanalysis2} in $t$, we have $v(R_c,t)={3R_c/(5-\omega)t}$ at $R_c$, and so we can solve for $f(t)$ to give\footnote{Note that Equation 11 of \cite{Weaver1977} contains a typo, and should read $R_c^3$, not $R_c^2$} }
\begin{equation}
    v(r,t) = \frac{11 - \omega}{25 - 5 \omega}\frac{R_c^3}{r^2 t} +  \frac{4 + \omega}{25 - 5\omega} \frac{r}{t}.
    \label{eqn:app_velocityequation}
\end{equation}

\rev{We adopt the solutions for an adiabatic wind in \cite{Weaver1977} in the region $R_1 \ll r \ll R_c$}
\begin{equation}
    v \sim \frac{v_w}{4}\left( \frac{15}{16}\right)^{3/2}\frac{R_1^2}{r^2}
    \label{eqn:app_velocity_rh}
\end{equation}
\rev{and}
\begin{equation}
    P \sim \frac{3}{4}\left( \frac{16}{15}\right)^{5/2}\frac{v_w \dot{M}_w}{\Omega R_1^2}
    \label{eqn:app_pressure_rh}
\end{equation}
\rev{where $v_w$ is the wind terminal velocity, and $\dot{M}_w$ is the wind mass loss rate. The wind luminosity is given by $L_w \equiv \frac{1}{2} \dot{M}_w v_w^2$. We also adopt a solid angle for a spherical segment $\Omega$, which equals $4 \pi$ for a full spherical wind bubble as in \cite{Weaver1977}.}

\rev{Noting the condition that $r \ll R_c$ as in \cite{Weaver1977}, we neglect the second term in Equation  \ref{eqn:app_velocityequation} and combine this with Equation \ref{eqn:app_velocity_rh} to give}
\begin{equation}
    R_1 = \left(\frac{44 - 4 \omega}{25 - 5\omega}\right)^{1/2}\left(\frac{16}{15}\right)^{3/4}\left(\frac{R_c^3}{v_w t}\right)^{1/2}
    \label{eqn:app_R1}
\end{equation}

\rev{Neglecting the kinetic energy in the wind bubble (since the solution is hydrostatic), the energy in the region $R_1 < r < R_c$ is given by}
\begin{equation}
    E_b = \frac{3}{2} \frac{\Omega}{3}\left(R_c^3 - R_1^3\right) P(t) \simeq \frac{\Omega}{2}R_c^3 P(t)
\end{equation}
\rev{Substituting Equations \ref{eqn:app_pressure_rh} and \ref{eqn:app_R1} into this, we arrive at}
\begin{equation}
    E_b = \left(\frac{5 - \omega}{11 - \omega}\right) L_w t
    \label{eqn:app_windenergypartition}
\end{equation}
\rev{which reduces to the \cite{Weaver1977} result for the case $\omega = 0$.}

\section{Comparison with Numerical Simulations}
\label{appendix:numericaltest}

\rev{In Appendix \ref{appendix:analyticenergypartition} we re-derive the fraction of energy from the wind stored in the hot wind bubble, versus the amount used to do work on the dense shell around it, as in \cite{Weaver1977} but for a power law density field. In this Section we verify the calculation using a numerical simulation.} 

\rev{The simulations were performed using \textsc{vh-1}\footnote{\url{http://wonka.physics.ncsu.edu/pub/VH-1/index.php}}, a freely available hydrodynamics code. We use a 1D spherically symmetric grid with radius 4 pc, subdivided into 2048 cells, numbered 0 to 2047, with identical radial thickness $\Delta x=403~$AU, and employ reflective boundary conditions. The initial gas state is a power law density field as in Equation \ref{eqn:powerlawdensity}, with $n_0 = 1000~$cm$^{-3}$, $r_0 = 1~$pc, and three simulations performed with $\omega = 0,1,2$ respectively. We set the density in cell 0 to the density in cell 1 to prevent a singularity at $r=0$. The temperature of the gas is set initially to 10 K everywhere, and $\gamma = 5/3$. No gravity, radiative transfer or radiative cooling is included.}

\rev{Throughout the simulation, we inject a constant source of mass $\dot{M}_w$, momentum $\dot{M}_w v_w$ and kinetic energy $L_w = \frac{1}{2}\dot{M}_w v_w^2$ into cell 0 directed outwards. We use $L_w=10^{36}~$erg/s and $\dot{M}_w=2.5\times10^{19}~$g/s, equivalent to a $\sim30~$\Msolar star. In addition to the inbuilt Courant condition in \textsc{vh-1}, we restrict the simulation timestep to be smaller than $\Delta x/(100~$km/s), in order to ensure that the first few timesteps before the wind bubble is established are not too long. We end the simulations at 200 kyr, before the edge of the swept-up material around the wind bubble encounters the edge of the simulation volume in any simulation.}

\begin{figure}
	\includegraphics[width=\columnwidth]{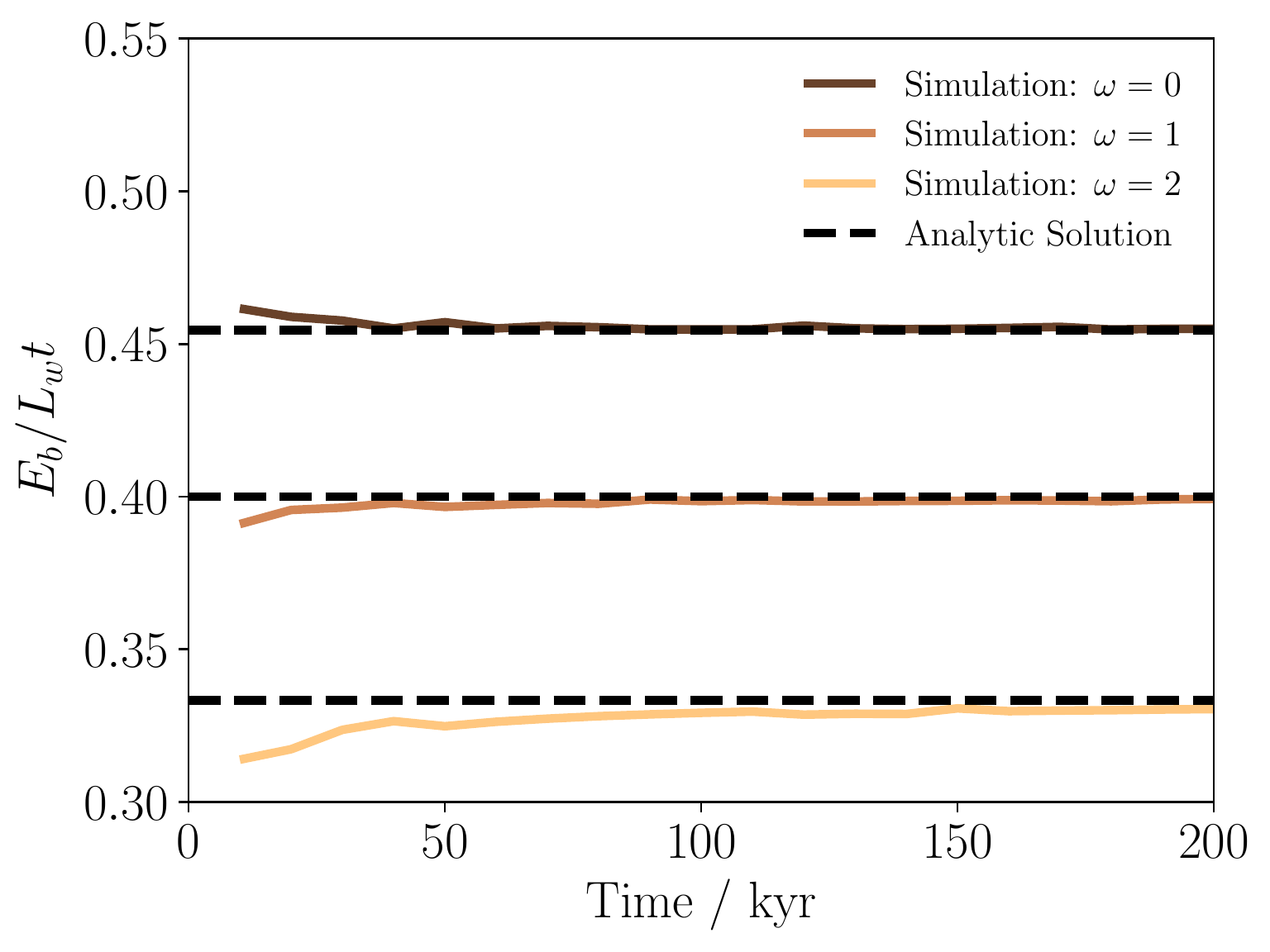}
	
    \caption{\rev{Comparison between Equation \ref{eqn:app_windenergypartition} and 1D numerical hydrodynamic simulations for the fraction of energy from stellar winds retained in the wind bubble, as opposed to the energy used to do work on the shell around the wind bubble. We show three simulations using initial power law density fields whose index is $\omega = 0,1,2$ as solid lines. Details of the simulations are given in Section \ref{appendix:numericaltest}. Dashed lines show the value of Equation \ref{eqn:app_windenergypartition} in each case. This figure shows that the energy fraction in the wind bubble converges to the analytic solution.}}
    \label{fig:app_energypartition}
\end{figure}

\rev{The energy in the bubble $E_b$ is defined to be the total energy in all cells above $10^5~$K, a threshold we select to distinguish between the hot wind bubble and the denser shell around it. We plot this quantity as a fraction of $L_w t$ against time in Figure \ref{fig:app_energypartition} for each simulation where $\omega = 0,1,2$, and overplot the value of Equation \ref{eqn:app_windenergypartition} in each case. Once the wind bubble is established, the results converge towards the analytic solution. The results converge faster where $\omega$ is smaller. We also note that the solution in Equation \ref{eqn:app_windenergypartition}, as with \cite{Weaver1977}, makes various approximations and is not an exact solution. A full analysis of the convergence to the analytic solution, including the role of sound waves (i.e. non-hydrostatic effects), is beyond the scope of this paper.}

\rev{We additionally plot the radial expansion of the wind bubble in these simulations against the analytic solution given in Equation \ref{eqn:rwt}, with $\omega=0$ reducing to the canonical \cite{Weaver1977} solution. For the simulations, we calculate the radius of the wind bubble as being the largest radius of any cell where the temperature is above $10^5~$K. There is good agreement in all cases. The radii converge around 100 kyr, although the expansion rates are still somewhat different.} 

\begin{figure}
	\includegraphics[width=\columnwidth]{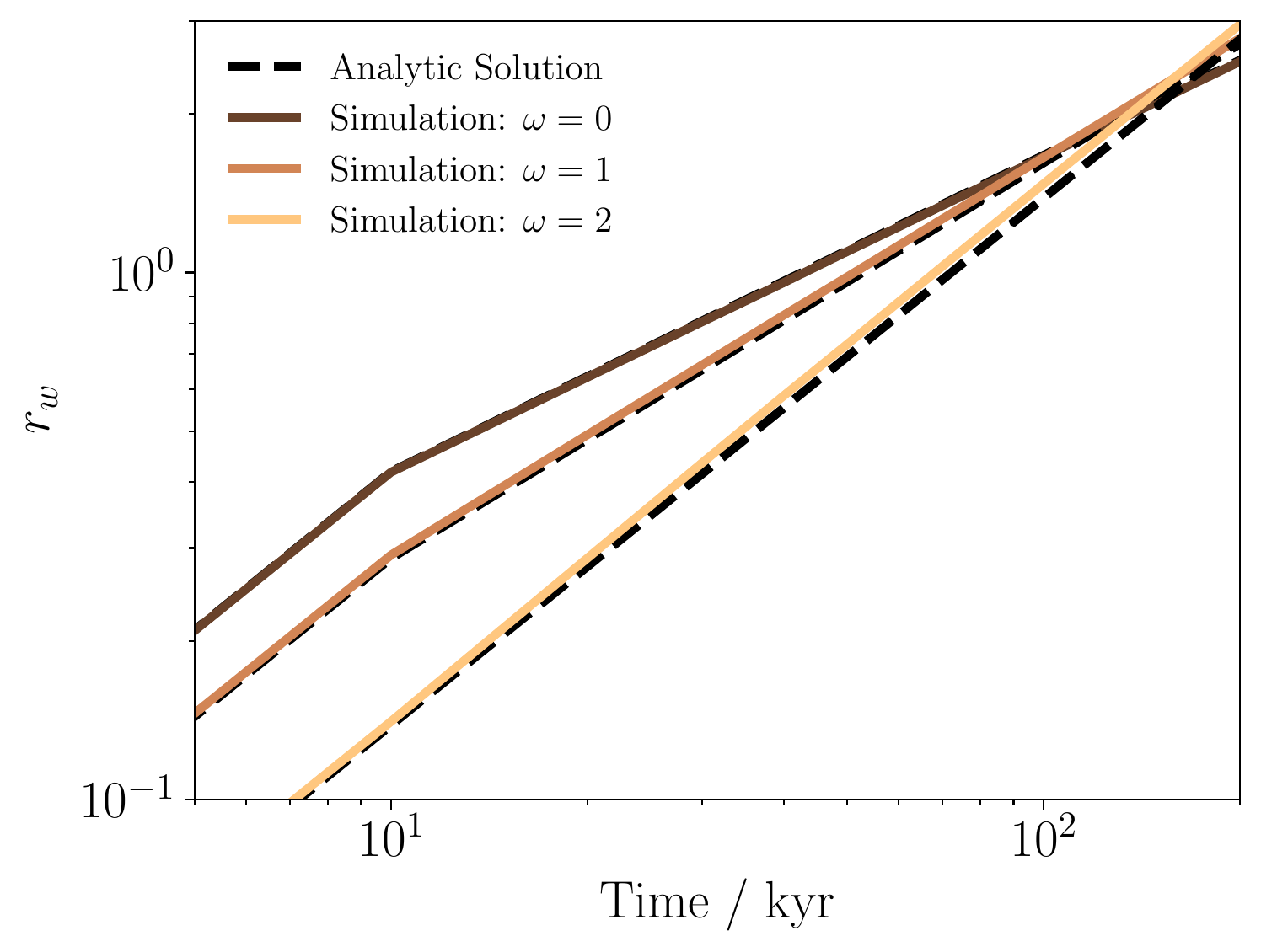}
    \caption{\rev{Comparison between Equation \ref{eqn:rwt} and 1D numerical hydrodynamic simulations for the radius of the wind bubble, using the same simulations and line styles as in Figure \ref{fig:app_energypartition}.}}
    \label{fig:app_radius}
\end{figure}

\section{Stellar Properties}
\label{appendix:starprops}

\begin{table*}

    \begin{center}
    \begin{tabular}{lllllllllllllllllllll}
    \multicolumn{1}{|c|}{\textbf{log$(M_{ini}$ / M$_{\odot})$}} & 
    \multicolumn{2}{|c|}{\textbf{log$(L_{w}$ / erg/s$)$}} & 
    \multicolumn{2}{|c|}{\textbf{log$(\dot{M}_w$ / M$_{\odot}$ / yr$)$}} &
    \multicolumn{2}{|c|}{\textbf{log$(L_{n}$ / erg/s$)$}} &
    \multicolumn{2}{|c|}{\textbf{log$(L_{i}$ / erg/s$)$}} &
    \multicolumn{2}{|c|}{\textbf{log$(Q_H / $s$^{-1})$}}  &
    \multicolumn{2}{|c|}{\textbf{log$(T_i / $K$)$}} \\
    \hline 
 120 & 37.51 & (36.8) & -5.01 & (-5.01) & 39.41 & (39.17) & 39.57 & (39.69) & 50.07 & (50.1) & 3.885 & (4.184) \\ 
115 & 37.47 & (36.75) & -5.058 & (-5.058) & 39.38 & (39.15) & 39.55 & (39.67) & 50.04 & (50.07) & 3.884 & (4.182) \\ 
110 & 37.42 & (36.71) & -5.107 & (-5.107) & 39.35 & (39.12) & 39.52 & (39.64) & 50.01 & (50.04) & 3.882 & (4.181) \\ 
105 & 37.38 & (36.67) & -5.156 & (-5.156) & 39.32 & (39.08) & 39.48 & (39.6) & 49.98 & (50.01) & 3.88 & (4.179) \\ 
100 & 37.33 & (36.62) & -5.205 & (-5.205) & 39.29 & (39.05) & 39.45 & (39.57) & 49.94 & (49.98) & 3.879 & (4.178) \\ 
95 & 37.28 & (36.57) & -5.253 & (-5.253) & 39.25 & (39.01) & 39.41 & (39.53) & 49.9 & (49.94) & 3.877 & (4.176) \\ 
90 & 37.23 & (36.52) & -5.301 & (-5.301) & 39.2 & (38.97) & 39.37 & (39.49) & 49.86 & (49.89) & 3.876 & (4.175) \\ 
85 & 37.17 & (36.46) & -5.35 & (-5.35) & 39.16 & (38.92) & 39.32 & (39.44) & 49.82 & (49.85) & 3.874 & (4.173) \\ 
80 & 37.1 & (36.39) & -5.428 & (-5.428) & 39.11 & (39.1) & 39.27 & (39.29) & 49.77 & (49.74) & 3.871 & (4.17) \\ 
75 & 37.02 & (36.32) & -5.507 & (-5.507) & 39.06 & (39.05) & 39.22 & (39.24) & 49.71 & (49.68) & 3.867 & (4.167) \\ 
70 & 36.94 & (36.24) & -5.586 & (-5.586) & 39.01 & (39.0) & 39.18 & (39.2) & 49.67 & (49.64) & 3.864 & (4.164) \\ 
65 & 36.85 & (36.16) & -5.664 & (-5.664) & 38.95 & (38.94) & 39.11 & (39.13) & 49.61 & (49.58) & 3.86 & (4.161) \\ 
60 & 36.77 & (36.07) & -5.743 & (-5.743) & 38.94 & (38.87) & 38.99 & (39.06) & 49.51 & (49.51) & 3.857 & (4.158) \\ 
55 & 36.63 & (35.94) & -5.882 & (-5.882) & 38.86 & (38.79) & 38.91 & (38.98) & 49.43 & (49.43) & 3.849 & (4.153) \\ 
50 & 36.48 & (35.81) & -6.021 & (-6.021) & 38.78 & (38.7) & 38.83 & (38.9) & 49.34 & (49.34) & 3.842 & (4.147) \\ 
45 & 36.33 & (35.66) & -6.161 & (-6.161) & 38.7 & (38.62) & 38.75 & (38.82) & 49.26 & (49.26) & 3.834 & (4.142) \\ 
40 & 36.18 & (35.51) & -6.301 & (-6.301) & 38.58 & (38.51) & 38.63 & (38.7) & 49.14 & (49.15) & 3.827 & (4.136) \\ 
35 & 35.94 & (35.29) & -6.525 & (-6.525) & 38.55 & (38.47) & 38.4 & (38.51) & 48.94 & (48.97) & 3.818 & (4.127) \\ 
30 & 35.66 & (35.02) & -6.793 & (-6.793) & 38.39 & (38.38) & 38.24 & (38.29) & 48.78 & (48.76) & 3.808 & (4.117) \\ 
25 & 35.3 & (34.68) & -7.127 & (-7.127) & 38.2 & (38.19) & 38.05 & (38.1) & 48.59 & (48.57) & 3.799 & (4.108) \\ 
20 & 34.79 & (34.2) & -7.598 & (-7.598) & 38.03 & (38.07) & 37.65 & (37.63) & 48.22 & (48.14) & 3.792 & (4.077) \\ 

    \end{tabular}
    \end{center}
    
\caption{\rev{Stellar properties used in this paper, for rotating stars (see Section \ref{overview:stellar_evolution_models}). Figures outside brackets are for stars at Solar metallicity ($Z=0.014$), with figures in brackets for stars at sub-Solar metallicity ($Z=0.02$). $M_{ini}$ is the initial mass of the star, $L_w$ is luminosity (i.e. mechanical power injection $\equiv 1/2 \dot{M}_{w} V_w^2$) of the wind, $\dot{M}_w$ is the mass loss rate, $L_{n}$ is the non-ionising radiative luminosity of the star (photons below 13.6 eV), $L_{i}$ is the total ionising luminosity, $Q_H$ is the hydrogen-ionising photon emission rate and $T_i$ is the temperature of the photionised gas around the star (assumed to be density-independent to good approximation).}}
\label{table:app_starprops}
\end{table*}

\begin{table*}

    \begin{center}
    \begin{tabular}{lllllllllllllllllllll}
    \multicolumn{1}{|c|}{\textbf{log$(M_{ini}$ / M$_{\odot})$}} & 
    \multicolumn{2}{|c|}{\textbf{log$(L_{w}$ / erg/s$)$}} & 
    \multicolumn{2}{|c|}{\textbf{log$(\dot{M}_w$ / M$_{\odot}$ / yr$)$}} &
    \multicolumn{2}{|c|}{\textbf{log$(L_{n}$ / erg/s$)$}} &
    \multicolumn{2}{|c|}{\textbf{log$(L_{i}$ / erg/s$)$}} &
    \multicolumn{2}{|c|}{\textbf{log$(Q_H / $s$^{-1})$}}  &
    \multicolumn{2}{|c|}{\textbf{log$(T_i / $K$)$}} \\
    \hline 
 120 & 37.5 & (36.77) & -5.035 & (-5.035) & 39.42 & (39.17) & 39.58 & (39.69) & 50.07 & (50.1) & 3.893 & (4.191) \\ 
115 & 37.46 & (36.73) & -5.083 & (-5.083) & 39.39 & (39.15) & 39.55 & (39.67) & 50.05 & (50.07) & 3.892 & (4.189) \\ 
110 & 37.41 & (36.69) & -5.13 & (-5.13) & 39.36 & (39.12) & 39.52 & (39.64) & 50.02 & (50.04) & 3.89 & (4.187) \\ 
105 & 37.37 & (36.65) & -5.177 & (-5.177) & 39.33 & (39.08) & 39.49 & (39.6) & 49.99 & (50.01) & 3.888 & (4.186) \\ 
100 & 37.32 & (36.6) & -5.224 & (-5.224) & 39.29 & (39.05) & 39.46 & (39.57) & 49.95 & (49.98) & 3.887 & (4.184) \\ 
95 & 37.27 & (36.55) & -5.271 & (-5.271) & 39.26 & (39.01) & 39.42 & (39.53) & 49.91 & (49.94) & 3.885 & (4.183) \\ 
90 & 37.22 & (36.51) & -5.318 & (-5.318) & 39.21 & (38.97) & 39.37 & (39.49) & 49.87 & (49.89) & 3.883 & (4.181) \\ 
85 & 37.17 & (36.45) & -5.366 & (-5.366) & 39.17 & (38.92) & 39.33 & (39.44) & 49.82 & (49.85) & 3.882 & (4.179) \\ 
80 & 37.09 & (36.38) & -5.443 & (-5.443) & 39.12 & (39.1) & 39.28 & (39.29) & 49.78 & (49.74) & 3.878 & (4.176) \\ 
75 & 37.02 & (36.31) & -5.519 & (-5.519) & 39.07 & (39.05) & 39.23 & (39.24) & 49.72 & (49.68) & 3.875 & (4.173) \\ 
70 & 36.94 & (36.24) & -5.596 & (-5.596) & 39.03 & (39.0) & 39.19 & (39.2) & 49.68 & (49.64) & 3.871 & (4.17) \\ 
65 & 36.86 & (36.16) & -5.673 & (-5.673) & 38.96 & (38.94) & 39.12 & (39.13) & 49.62 & (49.58) & 3.867 & (4.167) \\ 
60 & 36.77 & (36.07) & -5.75 & (-5.75) & 38.89 & (38.87) & 39.05 & (39.06) & 49.55 & (49.51) & 3.864 & (4.164) \\ 
55 & 36.64 & (35.95) & -5.886 & (-5.886) & 38.88 & (38.79) & 38.92 & (38.98) & 49.44 & (49.43) & 3.856 & (4.158) \\ 
50 & 36.5 & (35.81) & -6.022 & (-6.022) & 38.79 & (38.7) & 38.84 & (38.9) & 49.35 & (49.34) & 3.849 & (4.153) \\ 
45 & 36.35 & (35.67) & -6.157 & (-6.157) & 38.71 & (38.62) & 38.76 & (38.82) & 49.27 & (49.26) & 3.842 & (4.147) \\ 
40 & 36.2 & (35.53) & -6.293 & (-6.293) & 38.59 & (38.51) & 38.64 & (38.7) & 49.16 & (49.15) & 3.835 & (4.142) \\ 
35 & 35.96 & (35.3) & -6.517 & (-6.517) & 38.47 & (38.47) & 38.52 & (38.51) & 49.03 & (48.97) & 3.824 & (4.133) \\ 
30 & 35.68 & (35.04) & -6.783 & (-6.783) & 38.4 & (38.38) & 38.25 & (38.29) & 48.79 & (48.76) & 3.813 & (4.123) \\ 
25 & 35.33 & (34.71) & -7.111 & (-7.111) & 38.21 & (38.19) & 38.06 & (38.1) & 48.6 & (48.57) & 3.802 & (4.112) \\ 
20 & 34.82 & (34.24) & -7.593 & (-7.593) & 38.1 & (38.07) & 37.49 & (37.63) & 48.03 & (48.14) & 3.797 & (4.088) \\ 

    \end{tabular}
    \end{center}
    
\caption{\rev{As in Table \ref{table:app_starprops} but for non-rotating stars.}}
\label{table:app_starprops_nonrot}
\end{table*}

\rev{We give the stellar properties used in this paper in Table \ref{table:app_starprops} for rotating stars, and \ref{table:app_starprops_nonrot} for non-rotating stars. For simplicity, in this paper the values are taken to be constant throughout the star's lifetime, using the values at a stellar age of $10^5~$years. Stellar properties are taken from the Geneva stellar evolution models \citep{Ekstrom2012}, with atmospheres calculated using \textsc{Starburst99} \citep{Leitherer2014}. Wind luminosities are calculated by converting the surface escape velocity of the star into terminal velocity $V_w$ by using the prescription described in \cite{Gatto2017}. Further detailed descriptions of the techniques used to calculate the stellar models and plots of the evolutions of these properties at late times can be found in \cite{Geen2020}. Based on  Figure A1 of that paper, we justify our choice to use constant wind luminosities and photon emission rates by noting that these values do not change significantly for the first 1-2 Myr of the stars' lives.}


\bsp	
\label{lastpage}
\end{document}